\documentclass[twoside,twocolumn]{article}
\usepackage[super,sort&compress,comma]{natbib} 
\usepackage[version=3]{mhchem}
\usepackage[left=1.5cm, right=1.5cm, top=1.785cm, bottom=2.0cm]{geometry}
\usepackage{balance}
\usepackage{times,mathptmx}
\usepackage{sectsty}
\usepackage{soul}
\usepackage{graphicx}
\usepackage{kotex}
\usepackage{tabularx}
\usepackage{lastpage}
\usepackage[format=plain,justification=justified,singlelinecheck=false,font={stretch=1.125,small,sf},labelfont=bf,labelsep=space]{caption}
\usepackage{float}
\usepackage{fancyhdr}
\usepackage{fnpos}
\usepackage[english]{babel}
\addto{\captionsenglish}{%
  
}
\usepackage{array}
\usepackage{droidsans}
\usepackage{charter}
\usepackage[T1]{fontenc}
\usepackage[usenames,dvipsnames]{xcolor}
\usepackage{setspace}
\usepackage[compact]{titlesec}
\usepackage{hyperref}

\usepackage{epstopdf}
\AppendGraphicsExtensions{.tiff}

\usepackage{amsmath}
\usepackage{amsfonts}
\usepackage{siunitx}
\DeclareMathAlphabet{\mathcal}{OMS}{cmsy}{m}{n}

\definecolor{cream}{RGB}{222,217,201}

\usepackage{multirow}
\usepackage{parskip}

\begin{document}
\parskip 0pt plus 1pt

\pagestyle{fancy}
\fancypagestyle{plain}{

\fancyhead[C]{\includegraphics[width=18.5cm]{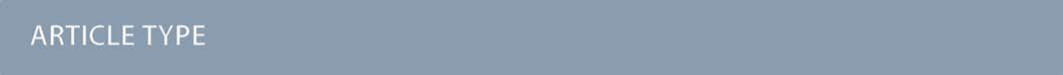}}
\fancyhead[L]{\hspace{0cm}\vspace{1.5cm}\includegraphics[height=30pt]{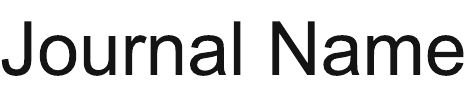}}
\fancyhead[R]{\hspace{0cm}\vspace{1.7cm}\includegraphics[height=55pt]{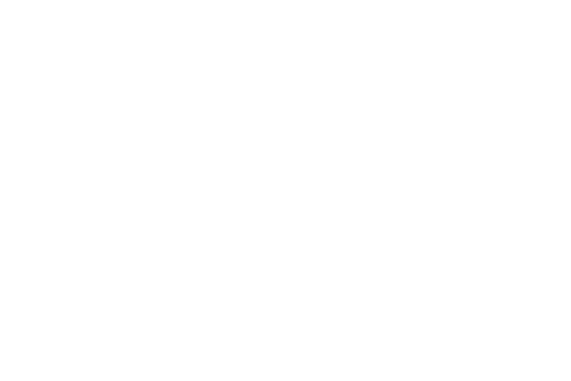}}
\renewcommand{\headrulewidth}{0pt}
}

\makeFNbottom
\makeatletter
\renewcommand\LARGE{\@setfontsize\LARGE{15pt}{17}}
\renewcommand\Large{\@setfontsize\Large{12pt}{14}}
\renewcommand\large{\@setfontsize\large{10pt}{12}}
\renewcommand\footnotesize{\@setfontsize\footnotesize{7pt}{10}}
\makeatother

\renewcommand{\thefootnote}{\fnsymbol{footnote}}
\renewcommand\footnoterule{\vspace*{1pt}%
\color{cream}\hrule width 3.5in height 0.4pt \color{black}\vspace*{5pt}} 
\setcounter{secnumdepth}{5}

\makeatletter 
\renewcommand\@biblabel[1]{#1}            
\renewcommand\@makefntext[1]%
{\noindent\makebox[0pt][r]{\@thefnmark\,}#1}
\makeatother 
\renewcommand{\figurename}{\small{Fig.}~}
\sectionfont{\sffamily\Large}
\subsectionfont{\normalsize}
\subsubsectionfont{\bf}
\setstretch{1.125} 
\setlength{\skip\footins}{0.8cm}
\setlength{\footnotesep}{0.25cm}
\setlength{\jot}{10pt}
\titlespacing*{\section}{0pt}{4pt}{4pt}
\titlespacing*{\subsection}{0pt}{15pt}{1pt}

\fancyfoot{}
\fancyfoot[LO,RE]{\vspace{-7.1pt}\includegraphics[height=9pt]{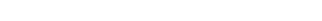}}
\fancyfoot[CO]{\vspace{-7.1pt}\hspace{13.2cm}\includegraphics{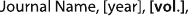}}
\fancyfoot[CE]{\vspace{-7.2pt}\hspace{-14.2cm}\includegraphics{head_foot/RF}}
\fancyfoot[RO]{\footnotesize{\sffamily{1--\pageref{LastPage} ~\textbar  \hspace{2pt}\thepage}}}
\fancyfoot[LE]{\footnotesize{\sffamily{\thepage~\textbar\hspace{2pt} 1--\pageref{LastPage}}}}
\fancyhead{}
\renewcommand{\headrulewidth}{0pt} 
\renewcommand{\footrulewidth}{0pt}
\setlength{\arrayrulewidth}{1pt}
\setlength{\columnsep}{6.5mm}
\setlength\bibsep{1pt}

\makeatletter 
\newlength{\figrulesep} 
\setlength{\figrulesep}{0.5\textfloatsep} 

\newcommand{\topfigrule}{\vspace*{-1pt}%
\noindent{\color{cream}\rule[-\figrulesep]{\columnwidth}{1.5pt}} }

\newcommand{\botfigrule}{\vspace*{-2pt}%
\noindent{\color{cream}\rule[\figrulesep]{\columnwidth}{1.5pt}} }

\newcommand{\dblfigrule}{\vspace*{-1pt}%
\noindent{\color{cream}\rule[-\figrulesep]{\textwidth}{1.5pt}} }

\makeatother

\twocolumn[
  \begin{@twocolumnfalse}
\vspace{3cm}
\sffamily
\begin{tabular}{m{4.5cm} p{13.5cm} }

\includegraphics{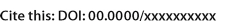} & \noindent\LARGE{\textbf{PIGNet: A physics-informed deep learning model toward generalized drug-target interaction predictions}} \\
\vspace{0.3cm} & \vspace{0.3cm} \\

 & \noindent\large{Seokhyun~Moon,\ddag\textsuperscript{a} Wonho~Zhung,\ddag\textsuperscript{a} Soojung~Yang,\ddag\textsuperscript{a}\textsection, Jaechang~Lim,\textsuperscript{b} and Woo~Youn~Kim$^\ast$\textsuperscript{abc}} \\

\includegraphics{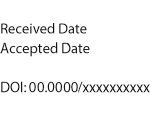} & \noindent\normalsize{%
Recently, deep neural network (DNN)-based drug-target interaction (DTI) models were highlighted for their high accuracy with affordable computational costs. Yet, the models' insufficient generalization remains a challenging problem in the practice of \textit{in-silico} drug discovery. We propose two key strategies to enhance generalization in the DTI model. The first is to predict the atom-atom pairwise interactions via physics-informed equations parameterized with neural networks and provides the total binding affinity of a protein-ligand complex as their sum. We further improved the model generalization by augmenting a broader range of binding poses and ligands to training data. We validated our model, PIGNet, in the comparative assessment of scoring functions (CASF) 2016, demonstrating the outperforming docking and screening powers than previous methods. Our physics-informing strategy also enables the interpretation of predicted affinities by visualizing the contribution of ligand substructures, providing insights for further ligand optimization.} \\

\end{tabular}

 \end{@twocolumnfalse} \vspace{0.6cm}

 ]

\renewcommand*\rmdefault{bch}\normalfont\upshape
\rmfamily
\section*{}
\vspace{-1cm}


\footnotetext{\textit{\textsuperscript{a}~Department of Chemistry, KAIST, 291 Daehak-ro, Yuseong-gu, Daejeon 34141, Republic of Korea. Email: wooyoun@kaist.ac.kr
}}
\footnotetext{\textit{\textsuperscript{b}~HITS incorporation, 124 Teheran-ro, Gangnam-gu, Seoul 06234, Republic of Korea. Email: jaechang@hits.ai}}
\footnotetext{\textit{\textsuperscript{c}~KI for Artificial Intelligence, KAIST, 291 Daehak-ro, Yuseong-gu, Daejeon 34141, Republic of Korea.}}

\footnotetext{\dag~Electronic Supplementary Information (ESI) available: the algorithm and implementation details of all the models used in the experiments, the details about the benchmark criteria and the physical interpretation result. See DOI: 00.0000/00000000.}

\footnotetext{\ddag~These authors contributed equally to this work.}
\footnotetext{$^\ast$~Corresponding author; E-mail:  wooyoun@kaist.ac.kr.}
\footnotetext{\textsection~\textit{Currently at Computational and Systems Biology, MIT, 77 Massachusetts Ave, Cambridge, MA}}






\section{Introduction}
\label{sec:introduction}
    Deep learning is a rapidly growing field of science.
The remarkable success of deep learning in various applications such as natural language processing, video games, and computer vision has raised expectations for similar success in other fields, leading to various applications of deep learning algorithms.
In particular, biomedical applications have become immediately one of the most active areas because they are not only socially influential but also scientifically challenging\cite{mamoshina2016applications,cao2018deep,zemouri2019deep}. 
Despite the great expectation, however, deep learning has not yet shown its highest potential in this field, due to low generalization issues caused by scarce and heavily imbalanced data\cite{wainberg2018deep,greener2021guide}.
Making a reliable prediction model for drug-target interaction, which is a key technology in the virtual screening of new drug candidates, is one such example\cite{hopkins2009predicting}.

For fast and yet reliable virtual screening, both high accuracy and low computational cost are critical. 
As an ideal virtual screening should be reliable yet fast, high accuracy and low computational cost are two essential factors of the virtual screening method.
However, docking methods\cite{trott2010autodock,Ruiz-carmona2014,Wang2002,jain2003surflex,Jones1997,Friesner2004,Venkatachalam2003,Korb2009,Allen2015a} as the most popular conventional approach are fast enough but insufficiently accurate\cite{Waszkowycz2011,Leach2006,Chen2015}, whereas more rigorous ones based on thermodynamic integration is computationally too expensive\cite{shirts2007alchemical,chipot2005can}.
While docking methods have been the most conventional approach due to their fast speed, they often show insufficient accuracy. More rigorous methods based on thermodynamic integration are more accurate but computationally too expensive.
Physics-based methods have an inherent limitation that low cost can only be achieved by losing accuracy as a trade-off. 
Such physics-based methods have an inherent limitation that low cost can only be achieved by losing their accuracy as a trade-off.
In contrast, a data-driven approach can improve prediction accuracy at no additional inference cost, just by learning with more data. This distinct feature of the data-driven approach has encouraged the active development of deep learning-based drug-target interaction (DTI) models that accomplish both high accuracy and low cost\cite{7863032,Ozturk2018,Thafar2019,Lipinski2019,Tsubaki2019,Lee2019,Zheng2020}. 

Among various deep learning-based models, the structure-based approach stands out for its accuracy; the spatial coordination of the protein and ligand is crucial in determining their interactions\cite{panday2019silico}.
Some of the promising studies utilize 3-dimensional convolutional neural networks (3D CNNs) \cite{Imrie2018,Stepniewska-dziubinska2018, Jimenez2018, Mohan, Ragoza2017,doi:10.1021/acs.jcim.9b00927, doi:10.1021/acsomega.9b01997,kwon2020ak,hassan2020rosenet,doi:10.1021/acs.jcim.0c01306},
graph neural networks (GNNs)\cite{Feinberg2018,Torng2019,Kim2019,doi:10.1021/acs.jcim.0c01306}, or feed-forward neural networks based on the atomic environment vectors\cite{meli2021learning}.
These state-of-the-art approaches had significantly improved the accuracy of DTI prediction compared to docking calculations.

Despite the advance of previous structure-based models, their limited generalization ability remains a challenging problem towards better performance. In particular, the deficiency in 3D structural data of the protein-ligand complexes could drive the models to excessively memorize the features in training data. Such models, being over-fitted to the training data, might fail to generalize in a broader context\cite{Hawkins2004}.
Several studies had suggested that deep learning-based models often learn the data-intrinsic bias instead of the underlying physics of the protein-ligand interaction as desired\cite{Chen2019,doi:10.1021/acs.jcim.9b00927, doi:10.1021/acs.jcim.0c00263}.
For instance, Chen \textit{et al.}\cite{Chen2019} reported an extremely high similarity in the performance of the receptor-ligand model and the ligand-only model - both trained with the DUD-E dataset - in terms of area under the ROC curve (AUC).
Such a similarity implies that the models might have learned to deduce the protein-ligand binding affinity only by looking at the ligand structures, regardless of whether or not the protein structures are included as inputs. Moreover, they showed that the such a memorization of wrong features can cause severe degradation in the performance for the proteins that have high structural variance from those in the training data.
The paper also reported that the 3D CNN and GNN models trained on the DUD-E  dataset had considerably underperformed when they were tested with the ChEMBL and MUV datasets\cite{Ragoza2017,Kim2019}.
Such an insufficient generalization of the DTI models can cause an increase in false positive rates in virtual screening scenarios, as the models would often fail to make correct predictions for unseen protein-ligand pairs.

In the field of physical applications of deep learning, the incorporation of appropriate physics as an inductive bias is a promising mean to improve the model generalization.
If a model is trained to obey certain physical principles, the model is expected to generalize to unseen data that is dictated by the same physics.
Several studies have indeed shown that the physics-informed models maintain their generalization ability for unseen data\cite{NIPS2019_9672,Pun,PhysRevLett.126.036401}.

In this regard, we propose two key strategies to enhance the generalization ability of DTI models. First, we introduce a novel physics-informed graph neural network, named PIGNet. It provides the binding affinity of a protein-ligand complex as a sum of atom-atom pairwise interactions, which are the combinations of the four energy components - van der Waals (VDW) interaction, hydrogen bond, metal-ligand interaction, and hydrophobic interaction. Each energy component is computed as an output of a physics model parameterized by deep neural networks, which learn the specific pattern of the interaction. 
This strategy can increase the generalization ability by allowing the model to dissect an unseen protein-ligand pair as combinations of commonly observed interactions between the protein and the ligand.
The detailed pattern of local interactions can render the model to learn the universal physics underlying the protein-ligand binding.
Moreover, as the model provides predictions for each atom-atom pair and each energy component, it is possible to analyze the contribution of individual molecular substructures to the binding affinity. This information can be used to modify drug candidates to further strengthen the binding affinity.  

Second, we leverage a data augmentation strategy. In practice, screening libraries include a variety of compounds where most of them do not appear in the training set.
Currently available experimental data on protein-ligand binding structures has very limited coverage on the structural diversity of all possible binding complexes.
A model trained with a set of experimental binding structures, which would only include the stable binding poses, may fail to distinguish the stable poses from non-stable poses in the inference set\cite{Chen2019}.
Therefore, we augmented our training data with computationally generated random binding poses of protein-ligand pairs to improve the model generalization\cite{doi:10.1021/acs.jcim.0c00263}.

To assess the generalization ability of the proposed model, we focused on the docking and the screening power of the CASF-2016 benchmark\cite{Su2019}. Previously, the DTI models had been evaluated in terms of the correlation between the predicted and the experimental binding affinities\cite{Ragoza2017,Mohan,Feinberg2018,Kim2019,Torng2019}. However, the high correlation does not automatically guarantee a good model generalization\cite{Chen2019}. A well-generalized model should be able to successfully identify the true binding pose that has minimum energy and correctly rank the best binding molecule.
The former criterion can be assessed in terms of docking power, while the latter one is related to screening power.
Examining a model for both tasks is essential to ensure the model's ability to generalize in real-world settings such as virtual high-throughput screening (vHTS).
We compared the benchmark results of PIGNet with traditional docking calculations and previous deep learning models and showed that our model significantly improved both docking power and screening power.

In addition to the improvement in the model performance, we show the interpretability of our model.
While interpreting the underlying chemistry of DTI prediction is an essential step of drug discovery, previous deep learning models that take a complete black box approach were not practical in that sense\cite{Pahikkala2015,10.5555/3304222.3304236}.
On the contrary, physics-based deep learning models can offer interpretability since several intermediate variables of the models have certain physical meanings\cite{zubatiuk2021machine}.
As our model predicts the interaction energy for each atom-atom pair, we can estimate the contribution of each ligand substructure in total binding free energy.
Such an interpretation can provide the guidelines for the practitioners regarding ligand optimization - modifying the less contributing moieties into stronger binding moieties can be an example.

\begin{figure*}[ht]
\includegraphics[width=\textwidth]{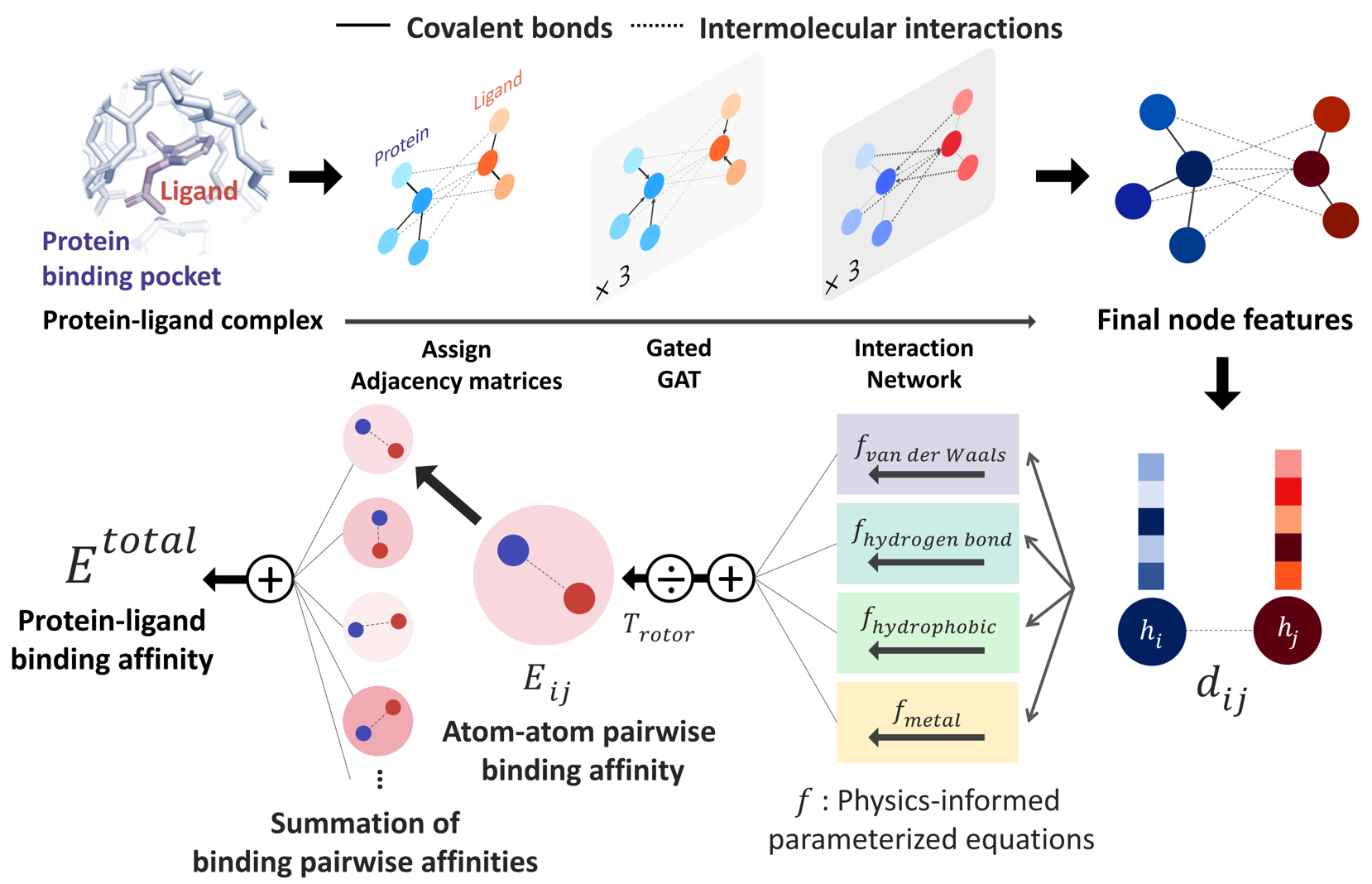}
    \caption{\textbf{Our model architecture.} 
    A protein-ligand complex is represented in a graph and adjacency matrices are assigned from the binding structure of the complex. Each node feature is updated through neural networks to carry the information of covalent bonds and intermolecular interactions. Given the distance and final node features of each atom pair, four energy components are calculated from the physics-informed parameterized equations. The total binding affinity is obtained as a sum of pairwise binding affinities, which is a sum of the four energy components divided by an entropy term.}
    \label{fig:architecture}
\end{figure*}

\section{Method}
\label{sec:Method}
\subsection{Related works}
\label{sec:Related works}
\subsubsection{Summary of previous deep learning-based DTI models}
The 3D CNN takes a 3D rectangular grid that represents the coordinate of atoms of a protein-ligand complex as an input\cite{Imrie2018,Stepniewska-dziubinska2018,Jimenez2018,Mohan,Ragoza2017,doi:10.1021/acs.jcim.9b00927,doi:10.1021/acsomega.9b01997,kwon2020ak,hassan2020rosenet,doi:10.1021/acs.jcim.0c01306}. The proposed 3D CNN models outperformed docking programs for the PDBbind and the DUD-E dataset in terms of Pearson's correlation coefficient and AUC, respectively.
Nevertheless, the high dimensionality of 3D rectangular representations and the absence of explicit representation of chemical interactions and bonds may put a limitation on 3D CNN models\cite{Karlov2020}.
One of the promising alternatives is a GNN, which represents structural information as molecular graphs\cite{ZHOU202057}. Each atom and chemical interaction (or bond) in a molecule is represented as a node and an edge in a graph, respectively.
Also, molecular graphs can incorporate 3D structural information by regarding an atom-atom pair as neighbors only if its pairwise Euclidean distance is within a certain threshold.
Moreover, graph representations are invariant to translations and rotations, unlike grid representations of 3D CNN.
Such advantages of graph representation over grid representation might have contributed to the state-of-the-art performance of GNNs in DTI predictions\cite{Feinberg2018,Torng2019,Kim2019,doi:10.1021/acs.jcim.0c01306,jiang2020drug}.

\subsubsection{Physics-informed neural networks}
Greydanus \textit{et al.}\cite{NIPS2019_9672} proposed the Hamiltonian neural network as an effective method to model the systems that follow Hamiltonian mechanics. They used deep neural networks to predict parameters in the Hamiltonian equation and showed better generalization than previous neural networks.
Pun \textit{et al.} \cite{Pun} proposed a physics-informed neural network for atomic potential modeling.
The model predicts the parameters of each type of interatomic potential energy, instead of directly predicting the total energy of the system. This strategy had improved the model generalization for simulations performed outside the bonding region.
In this work, with neural networks, we parameterize the equations that are derived from the physics of chemical interactions.

\subsection{Model architecture}
\label{sec:Model architecture}
PIGNet is a deep learning model that predicts binding free energy of a given protein-ligand complex structure (Fig. \ref{fig:architecture}). It takes a molecular graph, $G$, and the distances between the atom pairs, $d_{ij}$, of a protein-ligand complex as an input. Generally, a graph, $G$, can be defined as $(H,\ A)$, where $H$ is a set of node features and $A$ is an adjacency matrix. In an attributed graph, the $i^{th}$ node feature, $h_i$, is represented by a vector. 
Notably, our graph representation includes two adjacency matrices to discriminate the covalent bonds in each molecule and the intermolecular interactions between protein and ligand atoms. The details of the initial node features and the construction of the two adjacency matrices are explained in the Supplementary Information.

Our model consists of several units of gated graph attention network (gated GAT) and interaction network. The gated GAT and interaction network update each node feature via two adjacency matrices that correspond to covalent bonds and intermolecular interactions. During the node feature update, gated GAT and interaction network learn to convey the information of covalent bonds and intermolecular interactions, respectively. After several node feature updates, we calculate VDW interactions ($E^{vdw}$), hydrogen bond interactions ($E^{hbond}$), metal-ligand interactions ($E^{metal}$), and hydrophobic interactions ($E^{hydrophobic}$), by feeding the final node features into physics-informed parameterized equations. Specifically, for each energy component, the fully connected layers take a set of final node features as input and produce the parametric values of the physics-informed equation. We also consider the entropy loss from the protein-ligand binding by dividing total energy with rotor penalty ($T^{rotor}$). The total energy can be written as follows:
\begin{equation} \label{eq1}
    E^{total} = \frac {E^{vdw} + E^{hbond} + E^{metal}+E^{hydrophobic}}{T^{rotor}}.
\end{equation}

\setlength{\parindent}{0pt}
\subsubsection{Gated graph attention network (Gated GAT)}
The gated GAT updates a set of node features with respect to the adjacency matrix for covalent bonds. The attention mechanism aims to put different weights on the neighboring nodes regarding their importance. 
The attention coefficient, which implies the importance of the node, is calculated from the two nodes that are connected in a covalent bond and then normalized across the neighboring nodes. 
The purpose of the gate mechanism is to effectively deliver the information from the previous node features to the next node features. The extent of the contribution from the previous nodes is determined by a coefficient, which is obtained from the previous and new node features. We describe the details of gated GAT in the Supplementary Information.

\setlength{\parindent}{0pt}
\subsubsection{Interaction network}
The interaction network takes an updated set of node features from the gated GAT along with the adjacency matrix to generate the next set of node features. Unlike the gated GAT, the interaction network adopts an adjacency matrix featuring intermolecular interactions. The interaction network produces two different sets of embedded node features by multiplying the previous set of node features with two different learnable weights. Next, we apply max pooling to each set of embedded node features, obtaining two sets of interaction-embedded node features. The interaction embedded node features are then added to the embedded node features to generate the new node features. The final node features are obtained as a linear combination of the new and previous node features, where the linear combination is performed with a gated recurrent unit (GRU) \cite{Chung2014}. We describe the details of the interaction network in the Supplementary Information.

\subsection{Physics-informed parameterized function}
\label{sec:Physics-informed parameterized function}
PIGNet consists of four energy components - VDW interaction, hydrophobic interaction, hydrogen bonding, and metal-ligand interaction - and a rotor penalty. Energy component of an interaction between the $i^{th}$ node and the $j^{th}$ node is computed from two node features, $h_i$ and $h_j$. 
Since the node features contain the information of the two atoms and their interaction, the model can reasonably predict DTI.

The energy components and the rotor penalty are motivated from the empirical functions of AutoDock Vina\cite{trott2010autodock}. The total binding affinity is obtained as a weighted sum of energy components, where the weights are introduced to account for the difference between the calculated energies and the true free binding energies. PIGNet employs learnable parameters to find an optimal weight for each component, learning to account for the different types of protein-ligand interactions.  

Each energy component is calculated from $d_{ij}$ and $d'_{ij}$, which are the inter-atomic distance and the corrected sum of the VDW radii of the $i^{th}$ node and the $j^{th}$ node, respectively. $d'_{ij}$ can be represented as follows:
\begin{equation} \label{eq2}
    d'_{ij} = r_i + r_j + c\cdot b_{ij},
\end{equation}
where $r$ is the VDW radius of each node, which are taken from X-Score parameters\cite{Wang2002}. $b_{ij}$ is a correction term between the two nodes which is resulted from a fully connected layer that accepts two node features $h_i$ and $h_j$ as inputs.

\setlength{\parindent}{0pt}
\subsubsection{van der Waals (VDW) interaction}
We used 12-6 Lennard-Jones potential to calculate the VDW interaction term, $E^{vdw}$. We considered all protein and ligand atom pairs except for metal atoms whose VDW radii highly vary depending on the atom type. The total VDW energy is obtained as a sum of all possible atom-atom pairwise VDW energy contribution coefficients. $E^{vdw}$ can be described as follows:
\begin{equation} \label{eq3}
E^{vdw}=\sum_{i,j} c_{ij}\left[ {\left( \frac {d'_{ij}}{d_{ij}}\right) }^{12} - 2{\left( \frac {d'_{ij}}{d_{ij}}\right) }^6\right],
\end{equation}

where $c_{ij}$, predicted from a fully connected layer, indicates the minimum VDW interaction energy and renders each estimated energy component similar to the true energy component, in order to reflect the physical reality.  

\setlength{\parindent}{0pt}
\subsubsection{Hydrogen bond, Metal-ligand interaction, Hydrophobic interaction}
The pairwise energy contribution coefficients, $e_{ij}$, of hydrogen bond ($E^{hbond}$), metal-ligand interaction ($E^{metal}$), and hydrophobic interaction ($E^{hydrophobic}$) share the same expression as shown in equation (\ref{eq4}) with different coefficients, $c_1$, $c_2$, and a learnable scalar variable, $w$.
\begin{equation} \label{eq4}
e_{ij}=
\begin{cases}
 w & \text {if\quad} d_{ij}-d'_{ij} < c_1, \\
 w\left( \frac {d_{ij}-d'_{ij}-c_2}{c_1-c_2}\right) \qquad & \text {if\quad} c_1<d_{ij}-d'_{ij}<c_2, \\
 0 & \text {if\quad} d_{ij}-d'_{ij} > c_2
\end{cases}
\end{equation}

Here, $c_1$ and $c_2$ are set as -0.7 and 0.0 for hydrogen bonds and metal-ligand interactions, respectively, while the constants are set as 0.5 and 1.5 for hydrophobic interaction. We chose the same values of $c_1$ and $c_2$ for hydrogen bonds and metal-ligand interactions, since both originate from the electron donor-acceptor interactions. The total energy component is computed as a summation of all atom-atom pairwise energy contribution coefficients, as described in equation (\ref{eq5}):
\begin{equation} \label{eq5}
    E= \sum_{i,j} e_{ij}.
\end{equation}
We classified atoms into hydrogen bond acceptors, hydrogen bond donors, metal atoms, and hydrophobic atoms. Since hydrogen bonds appear between hydrogen bond donors and hydrogen bond acceptors, each atom that forms hydrogen bonds is selected by substructure matching of the general SMARTS\cite{Wunderlich2003} descriptors, which are summarized in the Table S2.
Metal atoms include $Mg$, $Ca$, $Mn$, $Fe$, $Co$, $Ni$, $Cu$, and $Zn$. Lastly, halogen atoms or carbon centers that are surrounded only by carbon or hydrogen atoms are classified as hydrophobic atoms\cite{Wang2002}.

\setlength{\parindent}{0pt}
\subsubsection{Rotor penalty}
The rotor penalty term, $T^{rotor}$, is intended to consider a loss of entropy as the binding pocket interrupts the free rotation of chemical bonds during protein-ligand binding. We assumed that the entropy loss is proportional to the number of the rotatable bonds of a ligand molecule. $T^{rotor}$ can be described as follows:
\begin{equation} \label{eq6}
    T^{rotor} = 1+C_{rotor}\times N_{rotor},
\end{equation}
where $N_{rotor}$ is the number of rotatable bonds and $C_{rotor}$ is a positive learnable scalar variable. We used RDKit software\cite{Landrum} to calculate $N_{rotor}$.

\setlength{\parindent}{0pt}
\subsection{Monte Carlo dropout (MCDO) and epistemic uncertainty}
\label{sec:Monte Carlo dropout (MCDO) and epistemic uncertainty}
A total of 30 models is ensembled during the test phase, with the same dropout ratio, 0.1, as the training phase. 
We obtained the predicted values by averaging individual predictions and interpreted the variances as epistemic uncertainties.
Here, we define PIGNet with and without MCDO as PIGNet (ensemble), PIGNet (single), respectively.

\begin{figure*}[ht]
\includegraphics[width=\textwidth]{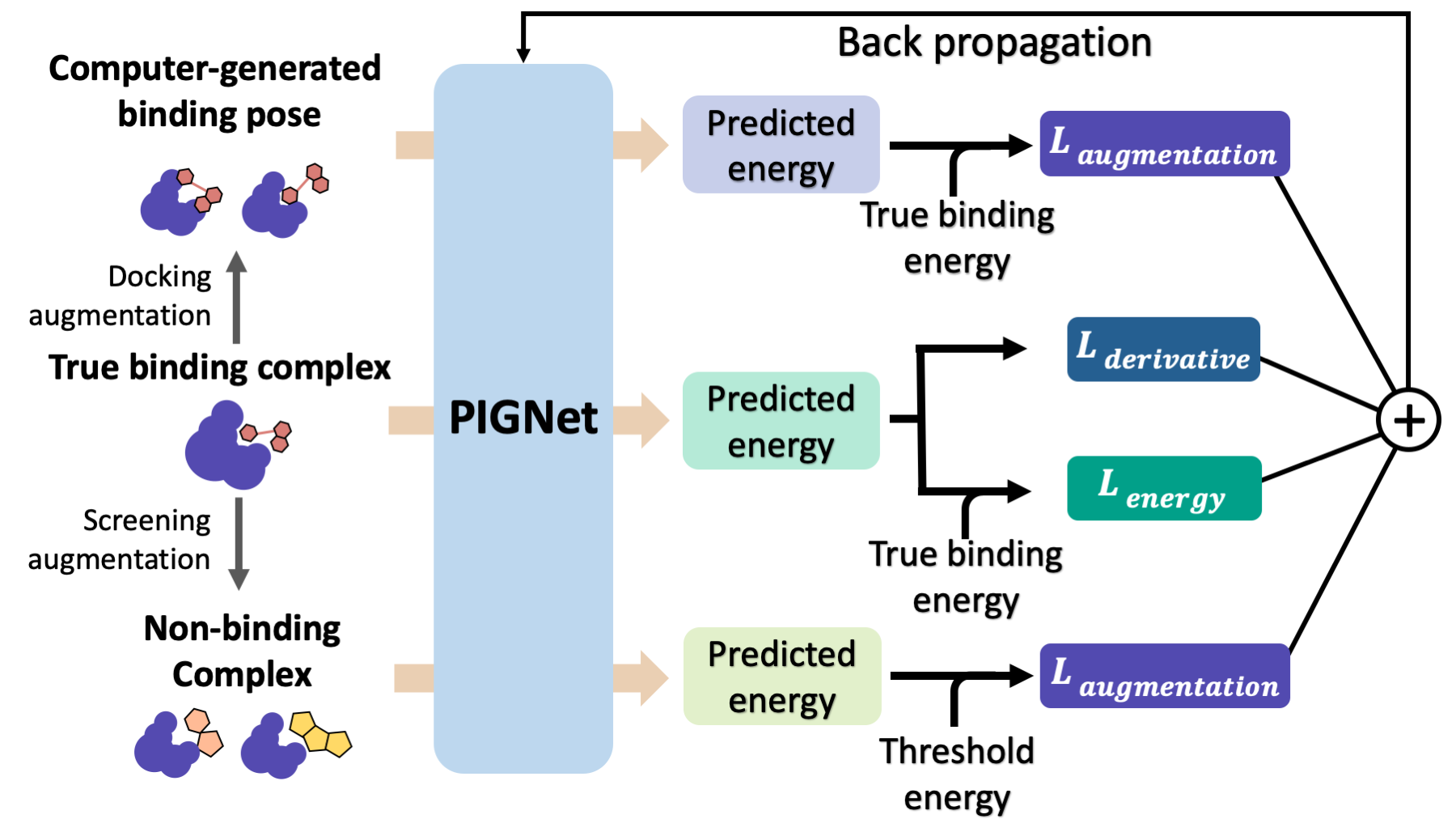}
    \caption{\textbf{The training scheme of PIGNet.} We use three types of data in model training - true binding complex, true binder ligand-protein pair in computer-generated binding pose, and non-binding decoy complex. PIGNet predicts binding free energy for each input. For a true binding complex, the model learns to predict its true binding energy. The model also learns to predict the energy of a computer-generated binding pose complex or a non-binding decoy complex in higher value than the true binding energy and threshold energy, respectively. Finally, PIGNet learns the proper correlation of ligand atom position and binding affinity by minimizing the derivative loss.}
    \label{fig:training scheme}
\end{figure*}

\subsection{Loss functions}
\label{sec:Loss functions}
The loss function of PIGNet consists of three components, $L_{energy}$, $L_{derivative}$, and $L_{augmentation}$ as in equation (\ref{eq7}): 
\begin{equation} \label{eq7}
    L_{total} = L_{energy} + L_{derivative} + L_{augmentation}.
\end{equation}

Fig. \ref{fig:training scheme} explains the overall training scheme of PIGNet based on the three loss functions. $L_{energy}$ is the mean squared error (MSE) loss between the predicted value from the model, $y_{pred}$, and the corresponding experimental binding free energy, $y_{true}$,
\begin{equation} \label{eq8}
    L_{energy}=\frac{1}{N_{train}}\sum_i \left( y_{pred,i}-y_{true,i}\right) ^2,
\end{equation}

where $N_{train}$ is a number of training data. Minimizing $L_{energy}$ enables the model to correctly predict the binding affinity of experimental 3D structures.
$L_{derivative}$ is composed of the first and the second derivative of the energy with respect to the atomic position. Minimizing $L_{derivative}$  intends the model to sensitively find relatively stable poses. $L_{augmentation}$ is the loss related to the data augmentation.

\setlength{\parindent}{0pt}
\subsubsection{Derivative loss}
The shape of the potential energy curve between the protein and ligand atoms has a huge impact on distinguishing the stable binding poses. The ligand atoms are located at the local minimum of the potential curve when the ligand binding is stable. Also, a potential energy curve in proper sharpness makes it easier to distinguish stable conformers from the others, as a small change in atomic positions would induce a large amount of energy deviation. Since a model trained with respect to $L_{energy}$ alone does not control the shape of the potential energy curve, it would be hard to distinguish whether or not a ligand is at a stable position. Accordingly, we guide the model with the derivative loss, $L_{derivative}$, to learn the proper shape of the pairwise potential energy curve, the width and the minimum energy position in particular.  

We can assume that the ligand atoms are located at the local minimum of the potential for the experimentally validated binding structures. Thus, we make the experimental structures as a local minimum by forcing the first derivative of the potential energy with respect to position to become zero. The sharpness of the potential energy curve was induced by increasing the second derivative. The derivative loss, $L_{derivative}$, is given as follows:
\begin{equation} \label{eq9}
    \hspace{-0.15cm} L_{derivative} = \sum_i \!\left[ \!\left( \frac{\partial E^{total}}{{\partial q_i}}\!\right) ^2 - \min\!\left( \!\left(\frac{\partial^2 E^{total}}{\partial q_i^2}\!\right) ,\ C_{der2}\!\right)\!\right],
\end{equation}

where $q_i$ is the position of the $i^{th}$ ligand atom. An excessively sharp potential energy curve may cause a problem in energy prediction by the immense deviation of energy from a small change in ligand atom positions. Therefore, we set its maximum value as $C_{der2}$, which is 20.0 in our model.

\setlength{\parindent}{0pt}
\subsubsection{Data augmentation loss}
Here, we constructed three different data augmentation-related loss functions; docking augmentation, random screening augmentation, and cross screening augmentation losses. 

\begin{itemize}
\item Docking augmentation loss \newline
The purpose of docking augmentation is to improve the model to distinguish the most stable binding poses from the others. We assume experimental binding structures from the PDBbind dataset\cite{Liu2017} as the most stable binding poses. Thus, the energy of experimental structures should be lower than the predicted energy of decoy structures that have different poses from true binding poses. The loss for docking augmentation, $L_{docking}$, can be written as follows:
\begin{equation} \label{eq10}
    L_{docking}=\sum_i \max\left( y_{exp,i}-y_{decoy,i},\ -1\right),
\end{equation}

where $y_{exp}$ is the energy of an experimental structure and $y_{decoy}$ is the predicted energy of a decoy structure. By minimizing $L_{docking}$, the model can predict $y_{decoy}$ larger than $y_{exp}+1$.  

\item Random screening augmentation loss \newline
In general, only a small fraction of molecules in a huge chemical space can bind to a specific target protein. Most molecules would have low binding affinity and high dissociation constant, $k_d$, with the target. From this nature, we assume that the dissociation constant of an arbitrary protein-ligand pair from the virtual screening library would be higher than $10^{-5}M$, as a criterion for hit identification is conventionally in micromolar ($10^{-6}M$) scale\cite{hughes2011principles}. Referring to the relationship between the binding free energy $\Delta G$ and the binding constant, $k_a$, which is reciprocal to $K_d$, we can set a threshold for $\Delta G$ of a protein-ligand pair as follows:
\begin{equation} \label{eq11}
    \Delta G \ge -1.36\log K_a=-6.8 \: \text{kcal/mol}.
\end{equation}

A model trained with random screening loss, $L_{random\_screening}$, and a non-binding random molecule-protein pair can sufficiently learn the chemical diversity. The model would predict the binding free energy of a random molecule with the target to a value higher than the threshold energy, $-6.8$. Thus, the loss for the random screening augmentation, $L_{random\_screening}$, can be written as follows:
\begin{equation} \label{eq12}
    L_{random\_screening}=\sum_i \max \left( -y_{random,i}-6.8, 0\right),
\end{equation}

where $y_{random}$ is the prediction energy of synthetic compounds from the IBS molecule library\cite{IBS}. The inaccuracy of a docking program is not problematic for the augmentation, as the binding energies of wrong binding poses are typically higher than the true binding energy.

\item Cross screening augmentation loss\newline
Another nature of protein-ligand binding is that if a ligand strongly binds to a specific target, the ligand is less likely to bind to other targets, because the different types of proteins have different binding pockets. We assumed that the true binders of the PDBbind dataset do not bind to the other proteins in the PDBbind dataset.

As in the random screening augmentation, training with non-binding ligands and protein pairs affect a model to learn chemical diversity. The loss for the cross screening augmentation, $L_{cross\_screening}$, can be written as follows: 
\begin{equation} \label{eq13}
    L_{cross\_screening}=\sum_i \max\left( -y_{cross,i}-6.8, 0\right),
\end{equation}
where $y_{cross}$ is the prediction energy of the cross binder. The same threshold for the binding free energy as in random screening augmentation is also used here.
\end{itemize}

\setlength{\parindent}{0pt}
\subsubsection{Total loss function}
The total loss, $L_{total}$, is the weighted sum of all the loss terms: $L_{energy}$, $L_{derivative}$, $L_{docking}$, $L_{random\_screening}$, and $L_{cross\_screening}$. The total loss can be written as follows:
\begin{equation} \label{eq14}
    \begin{split}
        L_{total}=& L_{energy}\\
        &+c_{derivative}L_{derivative}\\
        &+c_{docking}L_{docking}\\
        &+c_{random\_screening}L_{random\_screening}\\
        &+c_{cross\_screening}L_{cross\_screening},
    \end{split}
\end{equation}
where $c_{derivative}$, $c_{docking}$, $c_{random\_screening}$, and $c_{cross\_screening}$ are hyper-parameters which are set as 10.0, 10.0, 5.0, and 5.0, respectively.

\subsection{Baseline models}
\label{sec:Baseline models}
We constructed two baseline DNN models with the 3D CNN and 3D GNN architecture in comparison to PIGNet, namely a 3D CNN-based model and a 3D GNN-based model. For the 3D CNN-based model, we reimplemented the $K_{DEEP}$ model from Jiménez \textit{et al.}\cite{Jimenez2018} Our rebuilt 3D CNN-based model is identical to $K_{DEEP}$ 's, except we replaced the atom feature with those of PIGNet. We also constructed the 3D GNN-based model from PIGNet, but the model produces final outputs via fully connected layers instead of the physically modeled parametric equations.
\subsection{Dataset}
\label{sec:Dataset}
\subsubsection{Training dataset and data augmentation}
Our primary training set is the PDBbind 2019 refined set which provides qualified binding affinity data and corresponding structure of protein-ligand complexes deposited in the protein databank (PDB)\cite{Liu2017}. We eliminated the redundant samples in the test set - the core set of PDBbind 2016 - from the training set. We used 4,514 samples for training set and 265 samples for test set which were remained after the data processing. During the processing, the amino acid residues whose minimum distance between the ligand is greater than 5\si{\angstrom} are cropped to reduce the number of atoms in the protein pocket.

Additionally, we constructed three different data augmentations; docking augmentation, random screening augmentation, and cross screening augmentation. Smina (scoring and minimization with AutoDock Vina)\cite{Koes2013} was used for generating decoy structures. 
For the docking augmentation, we generated 292,518 decoy structures using the PDBbind 2016 dataset. 
For the random screening augmentation and the cross screening augmentation, we generated 831,885 complexes using the IBS molecules\cite{IBS} and 527,682 complexes based on the random cross binding, respectively. 
Any complexes in the test set are excluded during the augmentation.

\subsubsection{Benchmark dataset}
The CASF-2016 dataset is originated from the PDBbind core set complexes. 
After data processing, we used 283 samples for the scoring and ranking, 22,340 samples for the docking, and 1,612,867 samples for the screening task.
For the CSAR NRC-HiQ benchmark data, we removed the samples that are overlapping with the PDBbind data to avoid possible biases. We used 48 and 37 samples from the  CSAR NRC-HiQ dataset 1 and 2, respectively. 

\begin{table*}[htbp]
\scriptsize
\centering
{\renewcommand{\arraystretch}{1.3}
\begin{tabular}{l*{7}{c}}
\hline
\multicolumn{1}{c}{\multirow{3}{*}{Model}} & \multicolumn{5}{c}{CASF-2016} & \multicolumn{2}{c}{CSAR NRC-HiQ} \\ \cline{2-8} 
& Docking & \multicolumn{2}{c}{Screening} & Scoring & Ranking &$\quad\ \;$Set 1$\quad\ \;$&$\quad\ \;$Set 2$\quad\ \;$\\ \cline{2-8} 
& Success Rate & Average EF & Success Rate & $R$ & $\rho$ &$R$&$R$\\ \hline
AutoDock Vina\cite{trott2010autodock} & 84.6\% & 7.7 & 29.8\% & 0.604 & 0.528 &-&-\\ \hline
GlideScore-SP\cite{Friesner2004} & 84.6\% & 11.4 & 36.8\% & 0.513 & 0.419 &-&-\\ \hline
ChemPLP@GOLD\cite{Korb2009} & 83.2\% & 11.9 & 35.1\% & 0.614 & 0.633 &-&-\\ \hline
$K_{DEEP}$\cite{kwon2020ak} & 29.1\% & - & - & 0.701 & 0.528 &-&-\\ \hline
AK-Score (Single)\cite{kwon2020ak} & 34.9\% & - & - & 0.719 & 0.572 &-&-\\ \hline
AK-Score (Ensemble)\cite{kwon2020ak} & 36.0\% & - & - & 0.812 & 0.67 &-&-\\ \hline
AEScore \cite{meli2021learning} & 35.8\% & - & - & \textbf{0.830} & 0.640 &-&-\\ \hline
$\Delta$-AEScore \cite{meli2021learning} & 85.6\% & 6.16 & 19.3\% & 0.740 & 0.590 &-&-\\ \hline
3D CNN-based model & 48.2\% & 3.9 & 10.1\% & 0.687 & 0.580 &0.738&\textbf{0.804}\\ \hline
3D GNN-based model & 66.6\% & 10.2 & 28.5\% & 0.689 & 0.629 &0.588&0.687\\ \hline
PIGNet (Single) & 85.8\% & 18.5 & 50.0\% & 0.749 & 0.668 &\textbf{0.774}&0.799\\ \hline
PIGNet (Ensemble) & \textbf{87.0\%} & \textbf{19.6} & \textbf{55.4\%} & 0.761 & \textbf{0.682} &0.768&0.800\\ \hline
\end{tabular}}
\singlespace
\caption{
    Benchmark test results on the CASF-2016 and the CSAR NRC-HiQ dataset. $R$, $\rho$ indicate Pearson correlation coefficient and Spearman's rank correlation coefficient, respectively. 
    {Top 1 score was used for a docking success rate, and top 1\% rate was used for an average EF and a screening success rate.}
    $\Delta_{Vina}RF_{20}$\cite{Wang2017} was excluded from the comparison, as it was fine-tuned on the PDBbind 2017 data, which in fact includes $\sim50\%$ of data in the CASF-2016 test set.
    The highest values of each column are shown in bold
    \label{tab:main result}
}
\end{table*}

\section{Results and Discussions}
\label{sec:results and discussions}

\subsection{Assessment of the model performance and the generalization ability}
\label{sec:Assessment of the model performance and the generalization ability}

We assessed the model with the CASF-2016 benchmark dataset\cite{Su2019}, which is originated from the PDBbind 2016 core set. The CASF-2016 benchmark provides four different assessment tasks - scoring, ranking, docking, and screening - to evaluate DTI models in several aspects of virtual screening. The scoring power measures a linear correlation of predicted binding affinities and experimental values, calculated by a Pearson’s correlation coefficient R. The ranking power measures an ability of a model to correctly rank the binding affinities of true binders of the actual binding pose, calculated by a Spearman’s rank-correlation coefficient $\rho$.  
These two metrics are designed to assess the model’s ability upon the stable-and-precise binding structures.
On the other hand, the docking power and the screening power deal with the unnatural structures which are generated computationally. 
The docking power measures an ability of a model to find out the native binding pose of a ligand among computer-generated decoys, quantified as a success rate within the top $N$ candidates. The screening power measures the ability of a model to identify the specific binding ligand for a given target protein among a set of random molecules, quantified as a success rate and an enhancement factor (EF) within the top $\alpha$ percent of candidates. Detailed equations of each metric are summarized in the Supplementary Information. 

In vHTS schemes, a DTI model should identify the most stable binding pose and correctly rank the protein-ligand pairs by their binding affinities at the same time. Indeed, the ranking, docking, and screening powers would be optimal if the model accurately predicts the value of the binding affinity for every given complex, that is, what the scoring power targets to achieve. However, experimental analysis on the CASF-2016 benchmark shows that the high scoring power does not guarantee the high screening and docking powers\cite{Su2019}. We attribute this inconsistency to a limitation in the CASF-2016 scoring power benchmark - the scoring power itself cannot be a single criterion of a DTI model performance evaluation. Accordingly, we highlighted the models' docking and screening powers as indicators of model generalization. Along with the CASF-2016 dataset, we also measured the Pearson's correlation coefficient for the CSAR NRC-HiQ (2010) 1 and 2 benchmark sets\cite{Dunbar2011} to evaluate the model on the external datasets.

Table \ref{tab:main result} summarizes the performance of PIGNets, baseline models, and other published works for the CASF-2016 and the CSAR NRC-HiQ benchmarks. 
The reference scores of docking methods - AutoDock Vina\cite{trott2010autodock}, GlideScore-SP\cite{Friesner2004}, and ChemPLP@GOLD\cite{Korb2009} - were taken from Su \textit{et al.}\cite{Su2019}, which ranks the first in a docking success rate, screening success rate, and screening EF, respectively.
The performance of other deep learning approaches except $K_{DEEP}$\cite{Jimenez2018} was directly taken from their references. The scores of $K_{DEEP}$ were taken from Kwon \textit{et al.}\cite{kwon2020ak} since the docking power is not included in its original work.

PIGNet, both single and ensemble models, outperformed all other previous works in the CASF-2016 docking and screening powers. 
Our best model achieves a top 1 docking success rate of 87\%, a top 1\% screening success rate of 55.4\%, and a top 1\% average EF of 19.6.
The scoring and ranking power outperformed the docking methods while competitive with other deep learning-based approaches. The performance on the CSAR NRC-HiQ 1 and 2 benchmarks showed consistent results with the CASF-2016 scoring power benchmark.

For baseline models, the 3D GNN-based model showed better performance on ranking, docking, and screening powers than the 3D CNN-based model. The difference might lead from the lack of the chemical interaction information in the 3D CNN-based model, where the 3D GNN-based model implicitly has.
However, the 3D CNN-based model and the 3D GNN-based model fail to achieve high docking power and screening power. We attribute such low docking power to model over-fitting on the true-binding complex structures and binding affinities. The models have produced inaccurate binding affinities for the computer-generated decoy structures, which are primarily queried for the docking and screening power test. The low docking power then leads to the low screening power, as the most stable binding conformer needs to be identified in order to find the true binder. From these observations, we suspect that the performance reports of the previously introduced deep DTI models have been overoptimistic. In contrast, PIGNet consistently shows high performance across the four CASF-2016 metrics and the external CSAR NRC-HiQ benchmarks. Such results imply that our model is properly fitted to the training data, and also has learned the proper features - the underlying physics of protein-ligand binding patterns. Moreover, the results remind us that the scoring power cannot be a single criterion measuring the model performance. In the following section, we analyze how much each of our strategy had contributed to the result through ablation studies.

\subsection{Ablation study of two main strategies}
\label{sec:Ablation study of two main strategies}
We attribute the improvement of the model performance to two major strategies that have been utilized; the physics-informed parameterized functions introduced in the previous section and the data augmentation.
In this section, we carried out an ablation study to decouple the effects of the two strategies and summarized the results in Table \ref{tab:ablation study table}. 

\begin{table*}[ht!]
\centering
{\renewcommand{\arraystretch}{1.3}
\scriptsize
\begin{tabular}{lcccccc}
\hline
\multicolumn{1}{c}{\multirow{3}{*}{Model}} & \multirow{3}{*}{Use Data Augmentation?} & \multicolumn{5}{c}{CASF-2016} \\ \cline{3-7}
\multicolumn{1}{c}{} & & Docking & \multicolumn{2}{c}{Screening} & Scoring & Ranking \\ \cline{3-7} 
\multicolumn{1}{c}{} & & Success Rate & Average EF & Success Rate & $R$ & $\rho$ \\ \hline
\multirow{2}{*}{3D GNN-based model} & No & 29.9\% & 1.4 & 4.9\% & \textbf{0.772} & 0.604\\ \cline{2-7} 
 & Yes & \textbf{66.6\%} & \textbf{10.2} & \textbf{28.5\%} & 0.689 & \textbf{0.629}\\ \hline
\multirow{2}{*}{PIGNet (Single)} & No & 77.4\% & 6.6 & 24.6\% & \textbf{0.792} & \textbf{0.672} \\ \cline{2-7} 
 & Yes & \textbf{85.8\%} & \textbf{18.5} & \textbf{50.0\%} & 0.749 & 0.668 \\ \hline
\end{tabular}}
\singlespace
\caption{The CASF-2016 benchmark results for the 3D GNN-based model and PIGNet (Single) with and without using data augmentation. The highest values within the same model are shown in bold
\label{tab:ablation study table}
}
\end{table*}

\setlength{\parindent}{0pt}
\subsubsection{Effect of the physics-informed parametrized functions}
We can observe the effect of the physics-informed model by comparing the performances of the 3D GNN-based model and PIGNet, since a 3D GNN-based model is identical to PIGNet except the parametric equations.
As expected, the effect was not critical for the CASF-2016 scoring and ranking powers. However, the employment of the physics-informed model has resulted in a significant increase in docking and screening powers.
We can infer that the incorporation of the parametric equations has contributed to enhance the model generalization.
Incorporating a certain form of equations may impose an excessive inductive bias on the model, which can lead to the model under-fitting. However, it turns out to be unlikely from the comparable scoring powers of PIGNet and the 3D GNN-based model.
Especially, PIGNet without data augmentation still shows better docking power than the 3D GNN-based model trained with the augmented data.
Although adding a large number of augmented data improves the performance, the data augmentation strategy itself cannot entirely replace the generalization effect given by the physics-informed model. Instead, the data augmentation and physical modeling improve the model in a complementary manner, as we can see from the following section. 

\setlength{\parindent}{0pt}
\subsubsection{Effect of the DTI-adapted data augmentation strategy}
The PDBbind dataset is one of the most representative training datasets for the data-driven DTI models, providing both 3D binding structures and the binding affinities of the protein-ligand complexes\cite{Liu2017}. However, the PDBbind dataset is suspected to hold an intrinsic bias\cite{Chen2019}; its ligands have insufficient chemical diversity and only the binding structures in minimum energy poses are given. To expand the chemical space which the model learns, we additionally included 1,652,085 augmented samples in the training set. In particular, computationally generated structures, which happen to be unstable than the actual structures, are used for learning.
Table \ref{tab:ablation study table} clearly shows the effect of the DTI-adapted data augmentation strategy on the generalization ability.
The augmentation apparently improved the docking and screening power of both the 3D GNN-based model and PIGNet.
It shows the applicability of our data augmentation strategy for a variety of DNN-based DTI models. For benchmarks only containing the true binding complexes - scoring power, ranking power, and the CSAR NRC-HiQ - it was an expected result that the data augmentation did not improve the scores, because the model learns to accurately distinguish the decoy and true binding complexes from the augmented data and the corresponding losses.

\subsection{Interpretation of the physically modeled outputs}
\label{sec:Interpretation of the physically modeled outputs} 
\begin{figure*}[!ht]
\includegraphics[width=\textwidth]{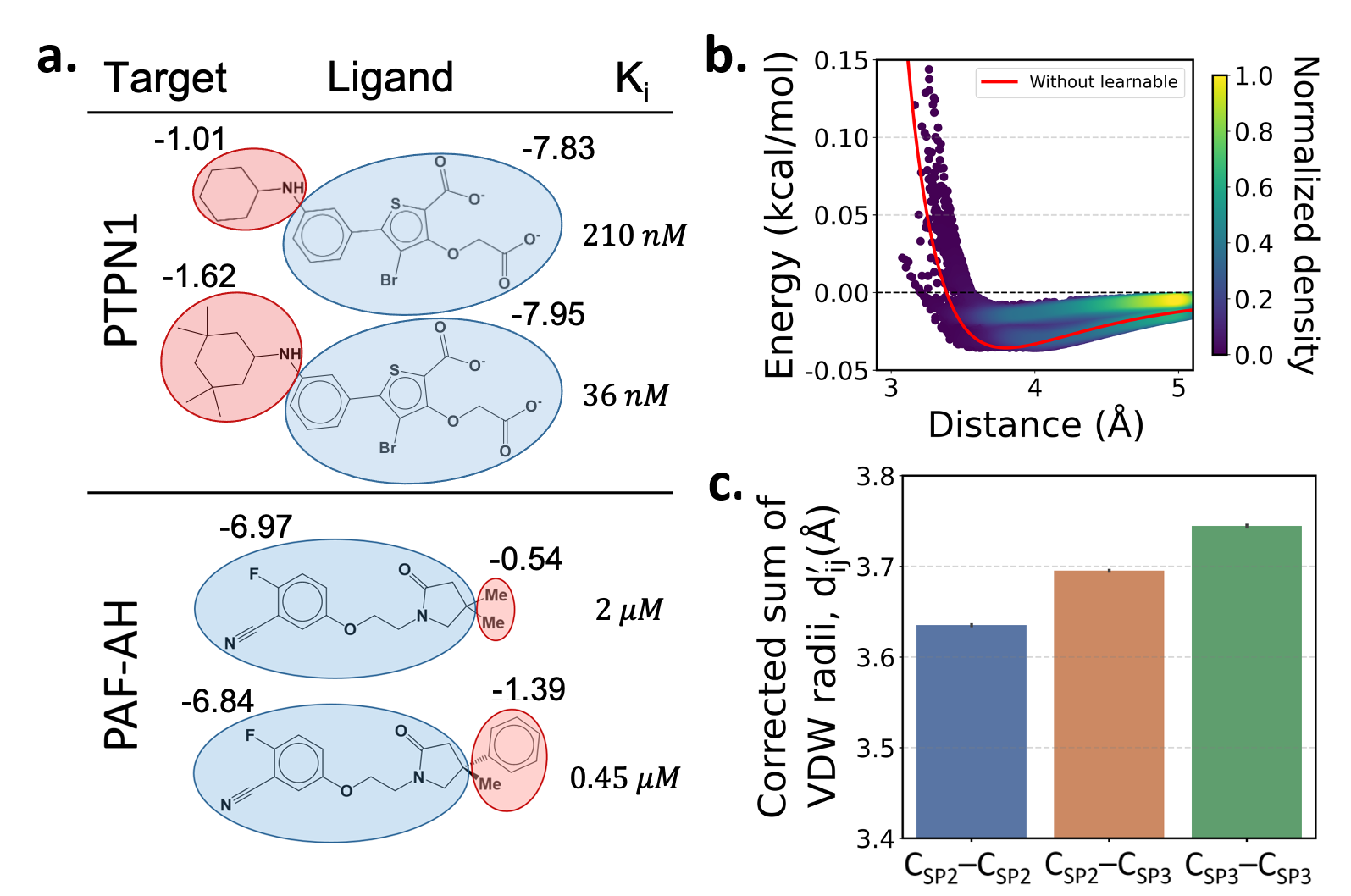}
    \caption{\textbf{Interpretation of the predicted outcomes.} \textbf{a.} Substructural analysis of ligands for two target proteins. Protein-tyrosine phosphatase non-receptor type 1 (PTPN1) and platelet activating factor acetylhydrolase (PAF-AH). The blue and red circles indicate common and different substructures, respectively, and the predicted energy contribution (unit: kcal/mol) of each substructure is annotated. The inhibitory constant, $K_i$, indicates how potent the ligand binds to the target protein. \textbf{b.} A distance-energy plot of carbon-carbon pairwise van der Waals (VDW) energy components in the test set. The red solid line illustrates the original distance-energy relation without any deviation induced by learnable parameters. The closer the color of a data point to yellow, the larger the number of corresponding carbon-carbon pairs. \textbf{c.} The average value of the corrected sum of VDW radii, $d'_{ij}$, corresponding to different carbon-carbon pair types. $C_{sp^2} - C_{sp^2}$, $C_{sp^2} - C_{sp^3}$, and $C_{sp^3} - C_{sp^3}$ pairs are compared. The results include 95\% confidence intervals.}
    \label{fig:interpretation}
\end{figure*}

One important advantage of our approach is the possibility of the atom-atom pairwise interpretation of DTI. To rationally design a drug for a specific target, knowing the dominant interaction of ligand binding is helpful. Since PIGNet computes atom-atom pairwise energy components, we can calculate the energy contribution of the substructures within a ligand. Here, we conduct a case study for two target proteins retrieved from the PDBbind dataset; protein-tyrosine phosphatase non-receptor type 1 (PTPN1) and platelet activating factor acetylhydrolase (PAF-AH). The result is illustrated in Fig. \ref{fig:interpretation}a, where two ligands for each protein are compared regarding the predicted substructural energy contributions and the inhibitory constant, $K_i$. 
Each ligand pair has high structural similarity, and only differ in red-circled moieties.
For PTPN1, the model predicts greater energy contribution for the tetramethyl cyclohexyl moiety than the cyclohexyl moiety.
Such a result is coherent to the experimental $K_{i}$ values.
For PAF-AH, the ligand with the phenyl group has a lower $K_i$ value than that of the ligand with the methyl group. The model predicts greater energy contribution of the phenyl group compared to the methyl group. 
For both proteins, the blue-circled common substructures are predicted to have similar energy contributions. This implies the predicted energy contribution of each substructure provides a physically meaningful interpretation, which can take further advantages to strengthen the total binding affinity towards the target protein during the ligand optimization.

Most docking programs manually assign different scoring functions to atom-atom pairs according to the predefined categories. This manual assignment would fall short when the binding pattern of the pair does not fit in the existing category.  
Instead of the handcrafted categorization, our model exploits neural networks to automatically differentiate the atom-atom pairs; the information of each pair's interaction is updated through the graph attention networks. We illustrate the deviation and its physical interpretation in Fig. \ref{fig:interpretation}b and \ref{fig:interpretation}c.  

Fig. \ref{fig:interpretation}b shows a distance-energy distribution plot of VDW component for carbon-carbon pairs within the test set. When trained with learnable parameters, predicted VDW interactions naturally deviate within the carbon-carbon pair, while without the learnable parameters the distance-energy plot follows a single solid line. With the aid of learnable parameters, our model might have learned a wider range of pairwise interactions in a data-driven manner. We also show the deviations in hydrophobic, hydrogen bond, and metal energy components in the Fig. S1.   

Fig. \ref{fig:interpretation}c shows that the naturally occurring deviations within the atom-atom pairs in our model are the consequences of learning sufficient physics information. The corrected sum of VDW radii, $d'_{ij}$, which contains a learnable parameter assigned to each atom-atom pair, deviated according to the carbon-carbon pair types. Since the interaction between the two carbon atoms would not be significantly affected by their hybridization, we speculate that the corrected sum of VDW radii of the pair would be dependent on the atom radii. The result shows an increasing tendency from the $C_{sp^2}-C_{sp^2}$ pair to the $C_{sp^3}-C_{sp^3}$ pair. Resonating with the speculation, larger the s-character of the carbon atoms, shorter was the corrected sum of VDW radii.

\begin{figure*}[!ht]
\centering
\includegraphics[width=1.0\textwidth]{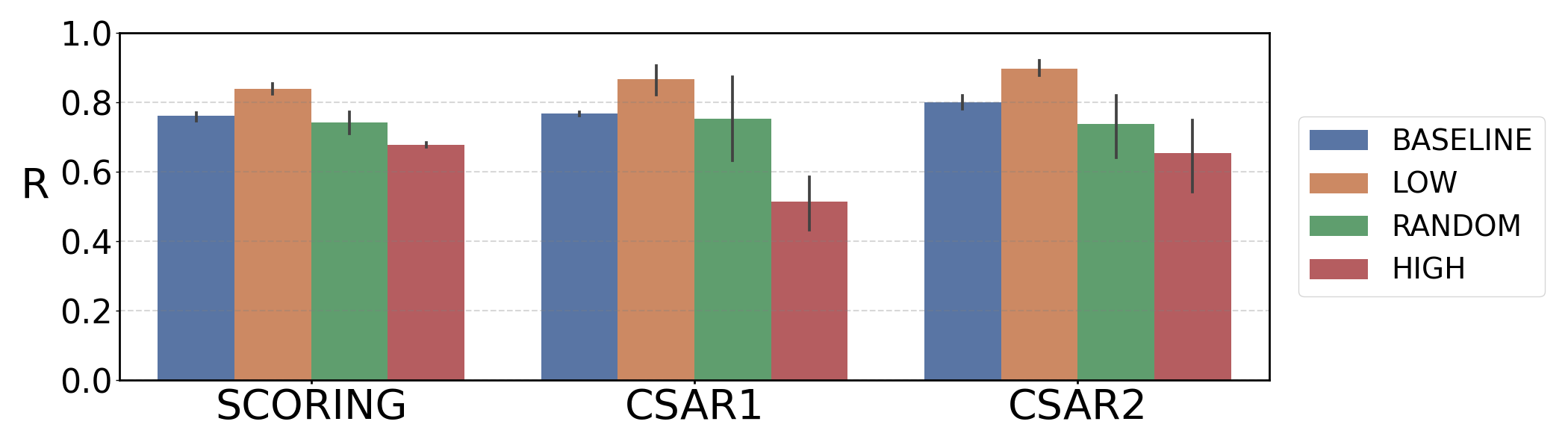}
    \caption{\textbf{Plot of the average Pearson's correlation coefficients, R, of the 5-fold PIGNet model, with or without the uncertainty estimator, on the datasets classified according to the total uncertainty.} PIGNet with the uncertainty estimator - \textbf{low}: the lowest third, \textbf{random}: the randomly selected one third, \textbf{high}: the highest third of the uncertainty distribution. PIGNet without Monte Carlo dropout - \textbf{baseline}:
    The scores of a single PIGNet model shown in the table \ref{tab:main result}.
    Error bars represent 95\% confidence intervals. PIGNet was tested at the $2,300^{th}$ training epoch with and without Monte Carlo dropout.
    }
    \label{fig:uncertainty figure}
\end{figure*}

\subsection{Epistemic uncertainty quantification of PIGNet}
\label{sec:epistemic uncertainty quantification of PIGNet}
For the reliable virtual screening, it is important to screen out the false positive binders and secure the true positives \cite{10.3389/fchem.2020.00343}. 
Unfortunately, most positive returns from docking programs turn out to be false positives \cite{Gobec2010}. DNN-based models may also have the same problem. In particular, the data-deficient nature of training DTI models might render the DNN models less fit to out-of-domain complexes \cite{Chen2019}, producing false positives.
One possible way to reduce the false positives is to use the uncertainty of the predictions and to filter unreliable positive predictions.
We employed a Monte Carlo dropout (MCDO), a practical Bayesian inference method utilizing dropout regularization, to estimate epistemic uncertainties which are originated from the model uncertainty\cite{pmlr-v48-gal16}.

We quantified prediction uncertainties for the samples in three datasets - CSAR NRC-HiQ 1 and 2, the CASF-2016 scoring power.
In Fig. \ref{fig:uncertainty figure}, the 'low', 'random', and 'high' batches in terms of the prediction uncertainties are in descending order in the value of Pearson's correlation, R. Such a result resonates with our expectation; the lower the uncertainty, the more probable the model would have correctly predicted the result. This result shows that the prediction uncertainties of our model can be properly quantified.
A previous study reported that the robust uncertainty quantification of model predictions can be an evidence of good generalization ability \cite{Scalia2020}. Thus, it might be possible to relate the high generalization ability of our model to the success in the uncertainty quantification.  

Comparing the R values of the 'random' and 'baseline' batches in Fig. \ref{fig:uncertainty figure} enables to evaluate the general performance of PIGNet with and without the uncertainty estimator. The result confirms that the addition of uncertainty estimator does not harm the model performance.
Furthermore, the comparison shows that uncertainty quantification can be used to filter out the false positives.

\parskip 0pt
\section{Conclusion}
\label{sec:conclusion}
In this work, we studied the inadequate generalization problem of deep learning models that are often encountered in real-world applications where data for training is very scarce and imbalanced. As an important practical example, we focused on drug target interaction (DTI) models for the fast and reliable virtual screening of drug candidates. The resulting model, named PIGNet, could achieve better generalization as well as higher accuracy compared to other deep learning models.
We attribute the success in our model to the following two strategies. The first one is to employ the physics-informed parameterized equations. The physics modeling acts as a proper inductive bias for the neural model, guiding the model to learn the underlying physics of the chemical interactions. We further improved the model performance by augmenting training data with protein-ligand complexes from the wider chemical and structural diversity. We analyzed the effects of the physics-informed model and the data augmentation through the ablation study and found that both contribute to the model generalization. 
These strategies can be readily adopted to other science fields where similar data problems are expected and the related physics is well-established. Such applications would include materials design, structural biology, and particle physics. The similar improvement in the generalization reported in the atomistic modeling of solid materials\cite{Pun} is one such example.

Our model can enjoy further practical advantages such as the physical interpretation of predicted DTI values and the reduction in false positives via uncertainty quantification. Obtaining binding free energy for every atom-atom pair opens up a possibility of further interpretation. This useful information can later be used to optimize drug candidates to attain better binding affinity by joining our model with generative models such as Imrie \textit{et al} \cite{imrie2020deep}. Also, we introduced an uncertainty estimator for DTI prediction models and evaluated the quality of estimation for PIGNet. As predictions in high uncertainty can possibly be false positives, the uncertainty quantification has practical benefits in virtual screening scenarios.

Still, our model has a room for improvement regarding the representation of solvation energy. In reality, proteins and ligands interact while surrounded by numerous water molecules, and the water molecules engender thermodynamic and structural effects on ligand binding \cite{Ulander2019}. As our model does not explicitly include a solvent energy component, we think introducing a solvent energy component to PIGNet will further improve the model.

\section*{Data Availability}
Training datasets can be preprocessed from the codes available at github: \url{https://github.com/jaechanglim/DTI_PDBbind.git}.

\section*{Conflicts of interest}
There are no conflicts to declare.

\section*{Author Contributions}
Conceptualization: J.L. and W.Y.K.; Methodology: S.M., W.Z., S.Y., and J.L.; Software, Investigation and Formal Analysis: S.M., W.Z., S.Y., and J.L.; Writing -- Original Draft: S.M., W.Z., S.Y., and J.L.; Writing -- Review \& Editing: S.M., W.Z., S.Y., J.L., and W.Y.K.; Supervision: W.Y.K.

\section*{Acknowledgements}
This work was supported by Basic Science Research Programs through the National Research Foundation of Korea (NRF) funded by the Ministry of Science and ICT (NRF-2017R1E1A1A01078109).




\bibliography{rsc} 

\providecommand*{\mcitethebibliography}{\thebibliography}
\csname @ifundefined\endcsname{endmcitethebibliography}
{\let\endmcitethebibliography\endthebibliography}{}
\begin{mcitethebibliography}{70}
\providecommand*{\natexlab}[1]{#1}
\providecommand*{\mciteSetBstSublistMode}[1]{}
\providecommand*{\mciteSetBstMaxWidthForm}[2]{}
\providecommand*{\mciteBstWouldAddEndPuncttrue}
  {\def\EndOfBibitem{\unskip.}}
\providecommand*{\mciteBstWouldAddEndPunctfalse}
  {\let\EndOfBibitem\relax}
\providecommand*{\mciteSetBstMidEndSepPunct}[3]{}
\providecommand*{\mciteSetBstSublistLabelBeginEnd}[3]{}
\providecommand*{\EndOfBibitem}{}
\mciteSetBstSublistMode{f}
\mciteSetBstMaxWidthForm{subitem}
{(\emph{\alph{mcitesubitemcount}})}
\mciteSetBstSublistLabelBeginEnd{\mcitemaxwidthsubitemform\space}
{\relax}{\relax}

\bibitem[Mamoshina \emph{et~al.}(2016)Mamoshina, Vieira, Putin, and
  Zhavoronkov]{mamoshina2016applications}
P.~Mamoshina, A.~Vieira, E.~Putin and A.~Zhavoronkov, \emph{Molecular
  pharmaceutics}, 2016, \textbf{13}, 1445--1454\relax
\mciteBstWouldAddEndPuncttrue
\mciteSetBstMidEndSepPunct{\mcitedefaultmidpunct}
{\mcitedefaultendpunct}{\mcitedefaultseppunct}\relax
\EndOfBibitem
\bibitem[Cao \emph{et~al.}(2018)Cao, Liu, Tan, Song, Shu, Li, Zhou, Bo, and
  Xie]{cao2018deep}
C.~Cao, F.~Liu, H.~Tan, D.~Song, W.~Shu, W.~Li, Y.~Zhou, X.~Bo and Z.~Xie,
  \emph{Genomics, proteomics \& bioinformatics}, 2018, \textbf{16},
  17--32\relax
\mciteBstWouldAddEndPuncttrue
\mciteSetBstMidEndSepPunct{\mcitedefaultmidpunct}
{\mcitedefaultendpunct}{\mcitedefaultseppunct}\relax
\EndOfBibitem
\bibitem[Zemouri \emph{et~al.}(2019)Zemouri, Zerhouni, and
  Racoceanu]{zemouri2019deep}
R.~Zemouri, N.~Zerhouni and D.~Racoceanu, \emph{Applied Sciences}, 2019,
  \textbf{9}, 1526\relax
\mciteBstWouldAddEndPuncttrue
\mciteSetBstMidEndSepPunct{\mcitedefaultmidpunct}
{\mcitedefaultendpunct}{\mcitedefaultseppunct}\relax
\EndOfBibitem
\bibitem[Wainberg \emph{et~al.}(2018)Wainberg, Merico, Delong, and
  Frey]{wainberg2018deep}
M.~Wainberg, D.~Merico, A.~Delong and B.~J. Frey, \emph{Nature biotechnology},
  2018, \textbf{36}, 829--838\relax
\mciteBstWouldAddEndPuncttrue
\mciteSetBstMidEndSepPunct{\mcitedefaultmidpunct}
{\mcitedefaultendpunct}{\mcitedefaultseppunct}\relax
\EndOfBibitem
\bibitem[Greener \emph{et~al.}(2021)Greener, Kandathil, Moffat, and
  Jones]{greener2021guide}
J.~G. Greener, S.~M. Kandathil, L.~Moffat and D.~T. Jones, \emph{Nature Reviews
  Molecular Cell Biology}, 2021,  1--16\relax
\mciteBstWouldAddEndPuncttrue
\mciteSetBstMidEndSepPunct{\mcitedefaultmidpunct}
{\mcitedefaultendpunct}{\mcitedefaultseppunct}\relax
\EndOfBibitem
\bibitem[Hopkins(2009)]{hopkins2009predicting}
A.~L. Hopkins, \emph{Nature}, 2009, \textbf{462}, 167--168\relax
\mciteBstWouldAddEndPuncttrue
\mciteSetBstMidEndSepPunct{\mcitedefaultmidpunct}
{\mcitedefaultendpunct}{\mcitedefaultseppunct}\relax
\EndOfBibitem
\bibitem[Trott and Olson(2010)]{trott2010autodock}
O.~Trott and A.~J. Olson, \emph{Journal of computational chemistry}, 2010,
  \textbf{31}, 455--461\relax
\mciteBstWouldAddEndPuncttrue
\mciteSetBstMidEndSepPunct{\mcitedefaultmidpunct}
{\mcitedefaultendpunct}{\mcitedefaultseppunct}\relax
\EndOfBibitem
\bibitem[Ruiz-Carmona \emph{et~al.}(2014)Ruiz-Carmona, Alvarez-Garcia, Foloppe,
  Garmendia-Doval, Juhos, Schmidtke, Barril, Hubbard, and
  Morley]{Ruiz-carmona2014}
S.~Ruiz-Carmona, D.~Alvarez-Garcia, N.~Foloppe, A.~B. Garmendia-Doval,
  S.~Juhos, P.~Schmidtke, X.~Barril, R.~E. Hubbard and S.~D. Morley, \emph{PLoS
  Computational Biology}, 2014, \textbf{10}, 1--7\relax
\mciteBstWouldAddEndPuncttrue
\mciteSetBstMidEndSepPunct{\mcitedefaultmidpunct}
{\mcitedefaultendpunct}{\mcitedefaultseppunct}\relax
\EndOfBibitem
\bibitem[Wang \emph{et~al.}(2002)Wang, Lai, and Wang]{Wang2002}
R.~Wang, L.~Lai and S.~Wang, \emph{Journal of Computer-Aided Molecular Design},
  2002, \textbf{16}, 11--26\relax
\mciteBstWouldAddEndPuncttrue
\mciteSetBstMidEndSepPunct{\mcitedefaultmidpunct}
{\mcitedefaultendpunct}{\mcitedefaultseppunct}\relax
\EndOfBibitem
\bibitem[Jain(2003)]{jain2003surflex}
A.~N. Jain, \emph{Journal of medicinal chemistry}, 2003, \textbf{46},
  499--511\relax
\mciteBstWouldAddEndPuncttrue
\mciteSetBstMidEndSepPunct{\mcitedefaultmidpunct}
{\mcitedefaultendpunct}{\mcitedefaultseppunct}\relax
\EndOfBibitem
\bibitem[Jones(1997)]{Jones1997}
G.~Jones, \emph{J. Mol. Biol.}, 1997, \textbf{267}, 727--748\relax
\mciteBstWouldAddEndPuncttrue
\mciteSetBstMidEndSepPunct{\mcitedefaultmidpunct}
{\mcitedefaultendpunct}{\mcitedefaultseppunct}\relax
\EndOfBibitem
\bibitem[Friesner \emph{et~al.}(2004)Friesner, Banks, Murphy, Halgren, Klicic,
  Mainz, Repasky, Knoll, Shelley, Perry,\emph{et~al.}]{Friesner2004}
R.~A. Friesner, J.~L. Banks, R.~B. Murphy, T.~A. Halgren, J.~J. Klicic, D.~T.
  Mainz, M.~P. Repasky, E.~H. Knoll, M.~Shelley, J.~K. Perry \emph{et~al.},
  \emph{Journal of medicinal chemistry}, 2004, \textbf{47}, 1739--1749\relax
\mciteBstWouldAddEndPuncttrue
\mciteSetBstMidEndSepPunct{\mcitedefaultmidpunct}
{\mcitedefaultendpunct}{\mcitedefaultseppunct}\relax
\EndOfBibitem
\bibitem[Venkatachalam \emph{et~al.}(2003)Venkatachalam, Jiang, Oldfield, and
  Waldman]{Venkatachalam2003}
C.~M. Venkatachalam, X.~Jiang, T.~Oldfield and M.~Waldman, \emph{Journal of
  Molecular Graphics and Modelling}, 2003, \textbf{21}, 289--307\relax
\mciteBstWouldAddEndPuncttrue
\mciteSetBstMidEndSepPunct{\mcitedefaultmidpunct}
{\mcitedefaultendpunct}{\mcitedefaultseppunct}\relax
\EndOfBibitem
\bibitem[Korb \emph{et~al.}(2009)Korb, St{\"u}tzle, and Exner]{Korb2009}
O.~Korb, T.~St{\"u}tzle and T.~E. Exner, \emph{Journal of Chemical Information
  and Modeling}, 2009, \textbf{49}, 84--96\relax
\mciteBstWouldAddEndPuncttrue
\mciteSetBstMidEndSepPunct{\mcitedefaultmidpunct}
{\mcitedefaultendpunct}{\mcitedefaultseppunct}\relax
\EndOfBibitem
\bibitem[Allen \emph{et~al.}(2015)Allen, Balius, Mukherjee, Brozell, Moustakas,
  Lang, Case, Kuntz, and Rizzo]{Allen2015a}
W.~J. Allen, T.~E. Balius, S.~Mukherjee, S.~R. Brozell, D.~T. Moustakas, P.~T.
  Lang, D.~A. Case, I.~D. Kuntz and R.~C. Rizzo, \emph{Journal of Computational
  Chemistry}, 2015, \textbf{36}, 1132--1156\relax
\mciteBstWouldAddEndPuncttrue
\mciteSetBstMidEndSepPunct{\mcitedefaultmidpunct}
{\mcitedefaultendpunct}{\mcitedefaultseppunct}\relax
\EndOfBibitem
\bibitem[Waszkowycz \emph{et~al.}(2011)Waszkowycz, Clark, and
  Gancia]{Waszkowycz2011}
B.~Waszkowycz, D.~E. Clark and E.~Gancia, \emph{WIREs Computational Molecular
  Science}, 2011, \textbf{1}, 229--259\relax
\mciteBstWouldAddEndPuncttrue
\mciteSetBstMidEndSepPunct{\mcitedefaultmidpunct}
{\mcitedefaultendpunct}{\mcitedefaultseppunct}\relax
\EndOfBibitem
\bibitem[Leach \emph{et~al.}(2006)Leach, Shoichet, and Peishoff]{Leach2006}
A.~R. Leach, B.~K. Shoichet and C.~E. Peishoff, \emph{Journal of Medicinal
  Chemistry}, 2006, \textbf{49}, 5851--5855\relax
\mciteBstWouldAddEndPuncttrue
\mciteSetBstMidEndSepPunct{\mcitedefaultmidpunct}
{\mcitedefaultendpunct}{\mcitedefaultseppunct}\relax
\EndOfBibitem
\bibitem[Chen(2015)]{Chen2015}
Y.~C. Chen, \emph{Trends in Pharmacological Sciences}, 2015, \textbf{36},
  78--95\relax
\mciteBstWouldAddEndPuncttrue
\mciteSetBstMidEndSepPunct{\mcitedefaultmidpunct}
{\mcitedefaultendpunct}{\mcitedefaultseppunct}\relax
\EndOfBibitem
\bibitem[Shirts \emph{et~al.}(2007)Shirts, Mobley, and
  Chodera]{shirts2007alchemical}
M.~R. Shirts, D.~L. Mobley and J.~D. Chodera, \emph{Annual reports in
  computational chemistry}, 2007, \textbf{3}, 41--59\relax
\mciteBstWouldAddEndPuncttrue
\mciteSetBstMidEndSepPunct{\mcitedefaultmidpunct}
{\mcitedefaultendpunct}{\mcitedefaultseppunct}\relax
\EndOfBibitem
\bibitem[Chipot \emph{et~al.}(2005)Chipot, Rozanska, and Dixit]{chipot2005can}
C.~Chipot, X.~Rozanska and S.~B. Dixit, \emph{Journal of computer-aided
  molecular design}, 2005, \textbf{19}, 765--770\relax
\mciteBstWouldAddEndPuncttrue
\mciteSetBstMidEndSepPunct{\mcitedefaultmidpunct}
{\mcitedefaultendpunct}{\mcitedefaultseppunct}\relax
\EndOfBibitem
\bibitem[Fitriawan \emph{et~al.}(2016)Fitriawan, Wasito, Syafiandini, Amien,
  and Yanuar]{7863032}
A.~Fitriawan, I.~Wasito, A.~F. Syafiandini, M.~Amien and A.~Yanuar, 2016
  International Conference on Computer, Control, Informatics and its
  Applications (IC3INA), 2016\relax
\mciteBstWouldAddEndPuncttrue
\mciteSetBstMidEndSepPunct{\mcitedefaultmidpunct}
{\mcitedefaultendpunct}{\mcitedefaultseppunct}\relax
\EndOfBibitem
\bibitem[{\"O}zt{\"u}rk \emph{et~al.}(2018){\"O}zt{\"u}rk, {\"O}zg{\"u}r, and
  Ozkirimli]{Ozturk2018}
H.~{\"O}zt{\"u}rk, A.~{\"O}zg{\"u}r and E.~Ozkirimli, \emph{Bioinformatics},
  2018, \textbf{34}, 821--829\relax
\mciteBstWouldAddEndPuncttrue
\mciteSetBstMidEndSepPunct{\mcitedefaultmidpunct}
{\mcitedefaultendpunct}{\mcitedefaultseppunct}\relax
\EndOfBibitem
\bibitem[Thafar \emph{et~al.}(2019)Thafar, Raies, Albaradei, Essack, and
  Bajic]{Thafar2019}
M.~Thafar, A.~B. Raies, S.~Albaradei, M.~Essack and V.~B. Bajic,
  \emph{Frontiers in Chemistry}, 2019, \textbf{7}, 782\relax
\mciteBstWouldAddEndPuncttrue
\mciteSetBstMidEndSepPunct{\mcitedefaultmidpunct}
{\mcitedefaultendpunct}{\mcitedefaultseppunct}\relax
\EndOfBibitem
\bibitem[Lipinski \emph{et~al.}(2019)Lipinski, Maltarollo, Oliveira, da~Silva,
  and Honorio]{Lipinski2019}
C.~F. Lipinski, V.~G. Maltarollo, P.~R. Oliveira, A.~B.~F. da~Silva and K.~M.
  Honorio, \emph{Frontiers in Robotics and AI}, 2019, \textbf{6}, 108\relax
\mciteBstWouldAddEndPuncttrue
\mciteSetBstMidEndSepPunct{\mcitedefaultmidpunct}
{\mcitedefaultendpunct}{\mcitedefaultseppunct}\relax
\EndOfBibitem
\bibitem[Tsubaki \emph{et~al.}(2019)Tsubaki, Tomii, and Sese]{Tsubaki2019}
M.~Tsubaki, K.~Tomii and J.~Sese, \emph{Bioinformatics}, 2019, \textbf{35},
  309--318\relax
\mciteBstWouldAddEndPuncttrue
\mciteSetBstMidEndSepPunct{\mcitedefaultmidpunct}
{\mcitedefaultendpunct}{\mcitedefaultseppunct}\relax
\EndOfBibitem
\bibitem[Lee \emph{et~al.}(2019)Lee, Keum, and Nam]{Lee2019}
I.~Lee, J.~Keum and H.~Nam, \emph{PLoS Computational Biology}, 2019,
  \textbf{15}, 1--21\relax
\mciteBstWouldAddEndPuncttrue
\mciteSetBstMidEndSepPunct{\mcitedefaultmidpunct}
{\mcitedefaultendpunct}{\mcitedefaultseppunct}\relax
\EndOfBibitem
\bibitem[Zheng \emph{et~al.}(2020)Zheng, Li, Chen, Xu, and Yang]{Zheng2020}
S.~Zheng, Y.~Li, S.~Chen, J.~Xu and Y.~Yang, \emph{Nature Machine
  Intelligence}, 2020, \textbf{2}, 134--140\relax
\mciteBstWouldAddEndPuncttrue
\mciteSetBstMidEndSepPunct{\mcitedefaultmidpunct}
{\mcitedefaultendpunct}{\mcitedefaultseppunct}\relax
\EndOfBibitem
\bibitem[Panday and Ghosh(2019)]{panday2019silico}
S.~K. Panday and I.~Ghosh, \emph{Structural Bioinformatics: Applications in
  Preclinical Drug Discovery Process}, 2019,  109--175\relax
\mciteBstWouldAddEndPuncttrue
\mciteSetBstMidEndSepPunct{\mcitedefaultmidpunct}
{\mcitedefaultendpunct}{\mcitedefaultseppunct}\relax
\EndOfBibitem
\bibitem[Imrie \emph{et~al.}(2018)Imrie, Bradley, {Van Der Schaar}, and
  Deane]{Imrie2018}
F.~Imrie, A.~R. Bradley, M.~{Van Der Schaar} and C.~M. Deane, \emph{Journal of
  Chemical Information and Modeling}, 2018, \textbf{58}, 2319--2330\relax
\mciteBstWouldAddEndPuncttrue
\mciteSetBstMidEndSepPunct{\mcitedefaultmidpunct}
{\mcitedefaultendpunct}{\mcitedefaultseppunct}\relax
\EndOfBibitem
\bibitem[Stepniewska-Dziubinska \emph{et~al.}(2018)Stepniewska-Dziubinska,
  Zielenkiewicz, and Siedlecki]{Stepniewska-dziubinska2018}
M.~M. Stepniewska-Dziubinska, P.~Zielenkiewicz and P.~Siedlecki,
  \emph{Bioinformatics}, 2018, \textbf{34}, 3666--3674\relax
\mciteBstWouldAddEndPuncttrue
\mciteSetBstMidEndSepPunct{\mcitedefaultmidpunct}
{\mcitedefaultendpunct}{\mcitedefaultseppunct}\relax
\EndOfBibitem
\bibitem[Jim{\'e}nez \emph{et~al.}(2018)Jim{\'e}nez, {\v{S}}kali{\v{c}},
  Mart{\'i}nez-Rosell, and {De Fabritiis}]{Jimenez2018}
J.~Jim{\'e}nez, M.~{\v{S}}kali{\v{c}}, G.~Mart{\'i}nez-Rosell and G.~{De
  Fabritiis}, \emph{Journal of Chemical Information and Modeling}, 2018,
  \textbf{58}, 287--296\relax
\mciteBstWouldAddEndPuncttrue
\mciteSetBstMidEndSepPunct{\mcitedefaultmidpunct}
{\mcitedefaultendpunct}{\mcitedefaultseppunct}\relax
\EndOfBibitem
\bibitem[Wallach \emph{et~al.}(2015)Wallach, Dzamba, and Heifets]{Mohan}
I.~Wallach, M.~Dzamba and A.~Heifets, \emph{arXiv}, 2015, preprint,
  arXiv:1510.02855, \url{https://arxiv.org/abs/1510.02855}\relax
\mciteBstWouldAddEndPuncttrue
\mciteSetBstMidEndSepPunct{\mcitedefaultmidpunct}
{\mcitedefaultendpunct}{\mcitedefaultseppunct}\relax
\EndOfBibitem
\bibitem[Ragoza \emph{et~al.}(2017)Ragoza, Hochuli, Idrobo, Sunseri, and
  Koes]{Ragoza2017}
M.~Ragoza, J.~Hochuli, E.~Idrobo, J.~Sunseri and D.~R. Koes, \emph{Journal of
  Chemical Information and Modeling}, 2017, \textbf{57}, 942--957\relax
\mciteBstWouldAddEndPuncttrue
\mciteSetBstMidEndSepPunct{\mcitedefaultmidpunct}
{\mcitedefaultendpunct}{\mcitedefaultseppunct}\relax
\EndOfBibitem
\bibitem[Morrone \emph{et~al.}(2020)Morrone, Weber, Huynh, Luo, and
  Cornell]{doi:10.1021/acs.jcim.9b00927}
J.~A. Morrone, J.~K. Weber, T.~Huynh, H.~Luo and W.~D. Cornell, \emph{Journal
  of Chemical Information and Modeling}, 2020, \textbf{60}, 4170--4179\relax
\mciteBstWouldAddEndPuncttrue
\mciteSetBstMidEndSepPunct{\mcitedefaultmidpunct}
{\mcitedefaultendpunct}{\mcitedefaultseppunct}\relax
\EndOfBibitem
\bibitem[Zheng \emph{et~al.}(2019)Zheng, Fan, and
  Mu]{doi:10.1021/acsomega.9b01997}
L.~Zheng, J.~Fan and Y.~Mu, \emph{ACS Omega}, 2019, \textbf{4},
  15956--15965\relax
\mciteBstWouldAddEndPuncttrue
\mciteSetBstMidEndSepPunct{\mcitedefaultmidpunct}
{\mcitedefaultendpunct}{\mcitedefaultseppunct}\relax
\EndOfBibitem
\bibitem[Kwon \emph{et~al.}(2020)Kwon, Shin, Ko, and Lee]{kwon2020ak}
Y.~Kwon, W.-H. Shin, J.~Ko and J.~Lee, \emph{International journal of molecular
  sciences}, 2020, \textbf{21}, 8424\relax
\mciteBstWouldAddEndPuncttrue
\mciteSetBstMidEndSepPunct{\mcitedefaultmidpunct}
{\mcitedefaultendpunct}{\mcitedefaultseppunct}\relax
\EndOfBibitem
\bibitem[Hassan-Harrirou \emph{et~al.}(2020)Hassan-Harrirou, Zhang, and
  Lemmin]{hassan2020rosenet}
H.~Hassan-Harrirou, C.~Zhang and T.~Lemmin, \emph{Journal of chemical
  information and modeling}, 2020, \textbf{60}, 2791--2802\relax
\mciteBstWouldAddEndPuncttrue
\mciteSetBstMidEndSepPunct{\mcitedefaultmidpunct}
{\mcitedefaultendpunct}{\mcitedefaultseppunct}\relax
\EndOfBibitem
\bibitem[Jones \emph{et~al.}(2021)Jones, Kim, Zhang, Zemla, Stevenson, Bennett,
  Kirshner, Wong, Lightstone, and Allen]{doi:10.1021/acs.jcim.0c01306}
D.~Jones, H.~Kim, X.~Zhang, A.~Zemla, G.~Stevenson, W.~F.~D. Bennett,
  D.~Kirshner, S.~E. Wong, F.~C. Lightstone and J.~E. Allen, \emph{Journal of
  Chemical Information and Modeling}, 2021, \textbf{61}, 1583--1592\relax
\mciteBstWouldAddEndPuncttrue
\mciteSetBstMidEndSepPunct{\mcitedefaultmidpunct}
{\mcitedefaultendpunct}{\mcitedefaultseppunct}\relax
\EndOfBibitem
\bibitem[Feinberg \emph{et~al.}(2018)Feinberg, Sur, Wu, Husic, Mai, Li, Sun,
  Yang, Ramsundar, and Pande]{Feinberg2018}
E.~N. Feinberg, D.~Sur, Z.~Wu, B.~E. Husic, H.~Mai, Y.~Li, S.~Sun, J.~Yang,
  B.~Ramsundar and V.~S. Pande, \emph{ACS Central Science}, 2018, \textbf{4},
  1520--1530\relax
\mciteBstWouldAddEndPuncttrue
\mciteSetBstMidEndSepPunct{\mcitedefaultmidpunct}
{\mcitedefaultendpunct}{\mcitedefaultseppunct}\relax
\EndOfBibitem
\bibitem[Torng and Altman(2019)]{Torng2019}
W.~Torng and R.~B. Altman, \emph{Journal of Chemical Information and Modeling},
  2019, \textbf{59}, 4131--4149\relax
\mciteBstWouldAddEndPuncttrue
\mciteSetBstMidEndSepPunct{\mcitedefaultmidpunct}
{\mcitedefaultendpunct}{\mcitedefaultseppunct}\relax
\EndOfBibitem
\bibitem[Lim \emph{et~al.}(2019)Lim, Ryu, Park, Choe, Ham, and Kim]{Kim2019}
J.~Lim, S.~Ryu, K.~Park, Y.~J. Choe, J.~Ham and W.~Y. Kim, \emph{Journal of
  Chemical Information and Modeling}, 2019, \textbf{59}, 3981--3988\relax
\mciteBstWouldAddEndPuncttrue
\mciteSetBstMidEndSepPunct{\mcitedefaultmidpunct}
{\mcitedefaultendpunct}{\mcitedefaultseppunct}\relax
\EndOfBibitem
\bibitem[Meli \emph{et~al.}(2021)Meli, Anighoro, Bodkin, Morris, and
  Biggin]{meli2021learning}
R.~Meli, A.~Anighoro, M.~J. Bodkin, G.~M. Morris and P.~C. Biggin,
  \emph{Journal of Cheminformatics}, 2021, \textbf{13}, 1--19\relax
\mciteBstWouldAddEndPuncttrue
\mciteSetBstMidEndSepPunct{\mcitedefaultmidpunct}
{\mcitedefaultendpunct}{\mcitedefaultseppunct}\relax
\EndOfBibitem
\bibitem[Hawkins(2004)]{Hawkins2004}
D.~M. Hawkins, \emph{Journal of Chemical Information and Computer Sciences},
  2004, \textbf{44}, 1--12\relax
\mciteBstWouldAddEndPuncttrue
\mciteSetBstMidEndSepPunct{\mcitedefaultmidpunct}
{\mcitedefaultendpunct}{\mcitedefaultseppunct}\relax
\EndOfBibitem
\bibitem[Chen \emph{et~al.}(2019)Chen, Cruz, Ramsey, Dickson, Duca, Hornak,
  Koes, and Kurtzman]{Chen2019}
L.~Chen, A.~Cruz, S.~Ramsey, C.~J. Dickson, J.~S. Duca, V.~Hornak, D.~R. Koes
  and T.~Kurtzman, \emph{PLoS one}, 2019, \textbf{14}, e0220113\relax
\mciteBstWouldAddEndPuncttrue
\mciteSetBstMidEndSepPunct{\mcitedefaultmidpunct}
{\mcitedefaultendpunct}{\mcitedefaultseppunct}\relax
\EndOfBibitem
\bibitem[Scantlebury \emph{et~al.}(2020)Scantlebury, Brown, Von~Delft, and
  Deane]{doi:10.1021/acs.jcim.0c00263}
J.~Scantlebury, N.~Brown, F.~Von~Delft and C.~M. Deane, \emph{Journal of
  Chemical Information and Modeling}, 2020, \textbf{60}, 3722--3730\relax
\mciteBstWouldAddEndPuncttrue
\mciteSetBstMidEndSepPunct{\mcitedefaultmidpunct}
{\mcitedefaultendpunct}{\mcitedefaultseppunct}\relax
\EndOfBibitem
\bibitem[Greydanus \emph{et~al.}(2019)Greydanus, Dzamba, and
  Yosinski]{NIPS2019_9672}
S.~Greydanus, M.~Dzamba and J.~Yosinski, \emph{Advances in Neural Information
  Processing Systems 32}, Curran Associates, Inc., 2019, pp. 15379--15389\relax
\mciteBstWouldAddEndPuncttrue
\mciteSetBstMidEndSepPunct{\mcitedefaultmidpunct}
{\mcitedefaultendpunct}{\mcitedefaultseppunct}\relax
\EndOfBibitem
\bibitem[Pun \emph{et~al.}(2019)Pun, Batra, Ramprasad, and Mishin]{Pun}
G.~P. Pun, R.~Batra, R.~Ramprasad and Y.~Mishin, \emph{Nature Communications},
  2019, \textbf{10}, 2339\relax
\mciteBstWouldAddEndPuncttrue
\mciteSetBstMidEndSepPunct{\mcitedefaultmidpunct}
{\mcitedefaultendpunct}{\mcitedefaultseppunct}\relax
\EndOfBibitem
\bibitem[Li \emph{et~al.}(2021)Li, Hoyer, Pederson, Sun, Cubuk, Riley, and
  Burke]{PhysRevLett.126.036401}
L.~Li, S.~Hoyer, R.~Pederson, R.~Sun, E.~D. Cubuk, P.~Riley and K.~Burke,
  \emph{Phys. Rev. Lett.}, 2021, \textbf{126}, 036401\relax
\mciteBstWouldAddEndPuncttrue
\mciteSetBstMidEndSepPunct{\mcitedefaultmidpunct}
{\mcitedefaultendpunct}{\mcitedefaultseppunct}\relax
\EndOfBibitem
\bibitem[Su \emph{et~al.}(2019)Su, Yang, Du, Feng, Liu, Li, and Wang]{Su2019}
M.~Su, Q.~Yang, Y.~Du, G.~Feng, Z.~Liu, Y.~Li and R.~Wang, \emph{Journal of
  Chemical Information and Modeling}, 2019, \textbf{59}, 895--913\relax
\mciteBstWouldAddEndPuncttrue
\mciteSetBstMidEndSepPunct{\mcitedefaultmidpunct}
{\mcitedefaultendpunct}{\mcitedefaultseppunct}\relax
\EndOfBibitem
\bibitem[Pahikkala \emph{et~al.}(2015)Pahikkala, Airola, Pietil{\"{a}},
  Shakyawar, Szwajda, Tang, and Aittokallio]{Pahikkala2015}
T.~Pahikkala, A.~Airola, S.~Pietil{\"{a}}, S.~Shakyawar, A.~Szwajda, J.~Tang
  and T.~Aittokallio, \emph{Briefings in bioinformatics}, 2015, \textbf{16},
  325--337\relax
\mciteBstWouldAddEndPuncttrue
\mciteSetBstMidEndSepPunct{\mcitedefaultmidpunct}
{\mcitedefaultendpunct}{\mcitedefaultseppunct}\relax
\EndOfBibitem
\bibitem[Gao \emph{et~al.}(2018)Gao, Fokoue, Luo, Iyengar, Dey, and
  Zhang]{10.5555/3304222.3304236}
K.~Y. Gao, A.~Fokoue, H.~Luo, A.~Iyengar, S.~Dey and P.~Zhang, Proceedings of
  the 27th International Joint Conference on Artificial Intelligence,
  2018\relax
\mciteBstWouldAddEndPuncttrue
\mciteSetBstMidEndSepPunct{\mcitedefaultmidpunct}
{\mcitedefaultendpunct}{\mcitedefaultseppunct}\relax
\EndOfBibitem
\bibitem[Zubatiuk \emph{et~al.}(2021)Zubatiuk, Nebgen, Lubbers, Smith,
  Zubatyuk, Zhou, Koh, Barros, Isayev, and Tretiak]{zubatiuk2021machine}
T.~Zubatiuk, B.~Nebgen, N.~Lubbers, J.~S. Smith, R.~Zubatyuk, G.~Zhou, C.~Koh,
  K.~Barros, O.~Isayev and S.~Tretiak, \emph{The Journal of Chemical Physics},
  2021, \textbf{154}, 244108\relax
\mciteBstWouldAddEndPuncttrue
\mciteSetBstMidEndSepPunct{\mcitedefaultmidpunct}
{\mcitedefaultendpunct}{\mcitedefaultseppunct}\relax
\EndOfBibitem
\bibitem[Karlov \emph{et~al.}(2020)Karlov, Sosnin, Fedorov, and
  Popov]{Karlov2020}
D.~S. Karlov, S.~Sosnin, M.~V. Fedorov and P.~Popov, \emph{ACS Omega}, 2020,
  \textbf{5}, 5150--5159\relax
\mciteBstWouldAddEndPuncttrue
\mciteSetBstMidEndSepPunct{\mcitedefaultmidpunct}
{\mcitedefaultendpunct}{\mcitedefaultseppunct}\relax
\EndOfBibitem
\bibitem[Zhou \emph{et~al.}(2020)Zhou, Cui, Hu, Zhang, Yang, Liu, Wang, Li, and
  Sun]{ZHOU202057}
J.~Zhou, G.~Cui, S.~Hu, Z.~Zhang, C.~Yang, Z.~Liu, L.~Wang, C.~Li and M.~Sun,
  \emph{AI Open}, 2020, \textbf{1}, 57--81\relax
\mciteBstWouldAddEndPuncttrue
\mciteSetBstMidEndSepPunct{\mcitedefaultmidpunct}
{\mcitedefaultendpunct}{\mcitedefaultseppunct}\relax
\EndOfBibitem
\bibitem[Jiang \emph{et~al.}(2020)Jiang, Li, Zhang, Wang, Wang, Yuan, and
  Wei]{jiang2020drug}
M.~Jiang, Z.~Li, S.~Zhang, S.~Wang, X.~Wang, Q.~Yuan and Z.~Wei, \emph{RSC
  Advances}, 2020, \textbf{10}, 20701--20712\relax
\mciteBstWouldAddEndPuncttrue
\mciteSetBstMidEndSepPunct{\mcitedefaultmidpunct}
{\mcitedefaultendpunct}{\mcitedefaultseppunct}\relax
\EndOfBibitem
\bibitem[Chung \emph{et~al.}(2014)Chung, Gulcehre, Cho, and Bengio]{Chung2014}
J.~Chung, C.~Gulcehre, K.~Cho and Y.~Bengio, \emph{arXiv}, 2014, preprint,
  arXiv:1412.3555, \url{https://arxiv.org/abs/1412.3555}\relax
\mciteBstWouldAddEndPuncttrue
\mciteSetBstMidEndSepPunct{\mcitedefaultmidpunct}
{\mcitedefaultendpunct}{\mcitedefaultseppunct}\relax
\EndOfBibitem
\bibitem[Wunderlich \emph{et~al.}(2003)Wunderlich, Wenisch, Falsafi, and
  Hoe]{Wunderlich2003}
R.~E. Wunderlich, T.~F. Wenisch, B.~Falsafi and J.~C. Hoe, \emph{Conference
  Proceedings - Annual International Symposium on Computer Architecture, ISCA},
  2003,  84--95\relax
\mciteBstWouldAddEndPuncttrue
\mciteSetBstMidEndSepPunct{\mcitedefaultmidpunct}
{\mcitedefaultendpunct}{\mcitedefaultseppunct}\relax
\EndOfBibitem
\bibitem[Lan()]{Landrum}
\emph{{RDKit: Open-source cheminformatics}}, \url{http://www.rdkit.org}\relax
\mciteBstWouldAddEndPuncttrue
\mciteSetBstMidEndSepPunct{\mcitedefaultmidpunct}
{\mcitedefaultendpunct}{\mcitedefaultseppunct}\relax
\EndOfBibitem
\bibitem[Liu \emph{et~al.}(2017)Liu, Su, Han, Liu, Yang, Li, and Wang]{Liu2017}
Z.~Liu, M.~Su, L.~Han, J.~Liu, Q.~Yang, Y.~Li and R.~Wang, \emph{Accounts of
  Chemical Research}, 2017, \textbf{50}, 302--309\relax
\mciteBstWouldAddEndPuncttrue
\mciteSetBstMidEndSepPunct{\mcitedefaultmidpunct}
{\mcitedefaultendpunct}{\mcitedefaultseppunct}\relax
\EndOfBibitem
\bibitem[Hughes \emph{et~al.}(2011)Hughes, Rees, Kalindjian, and
  Philpott]{hughes2011principles}
J.~P. Hughes, S.~Rees, S.~B. Kalindjian and K.~L. Philpott, \emph{British
  journal of pharmacology}, 2011, \textbf{162}, 1239--1249\relax
\mciteBstWouldAddEndPuncttrue
\mciteSetBstMidEndSepPunct{\mcitedefaultmidpunct}
{\mcitedefaultendpunct}{\mcitedefaultseppunct}\relax
\EndOfBibitem
\bibitem[IBS()]{IBS}
\emph{{InterBioScreen Ltd}}, \url{http://www.ibscreen.com}\relax
\mciteBstWouldAddEndPuncttrue
\mciteSetBstMidEndSepPunct{\mcitedefaultmidpunct}
{\mcitedefaultendpunct}{\mcitedefaultseppunct}\relax
\EndOfBibitem
\bibitem[Koes \emph{et~al.}(2013)Koes, Baumgartner, and Camacho]{Koes2013}
D.~R. Koes, M.~P. Baumgartner and C.~J. Camacho, \emph{Journal of Chemical
  Information and Modeling}, 2013, \textbf{53}, 1893--1904\relax
\mciteBstWouldAddEndPuncttrue
\mciteSetBstMidEndSepPunct{\mcitedefaultmidpunct}
{\mcitedefaultendpunct}{\mcitedefaultseppunct}\relax
\EndOfBibitem
\bibitem[Wang and Zhang(2017)]{Wang2017}
C.~Wang and Y.~Zhang, \emph{Journal of Computational Chemistry}, 2017,
  \textbf{38}, 169--177\relax
\mciteBstWouldAddEndPuncttrue
\mciteSetBstMidEndSepPunct{\mcitedefaultmidpunct}
{\mcitedefaultendpunct}{\mcitedefaultseppunct}\relax
\EndOfBibitem
\bibitem[Jr. \emph{et~al.}(2011)Jr., Smith, Yang, Ung, Lexa, Khazanov, Stuckey,
  Wang, and Carlson]{Dunbar2011}
J.~B.~D. Jr., R.~D. Smith, C.-Y. Yang, P.~M.-U. Ung, K.~W. Lexa, N.~A.
  Khazanov, J.~A. Stuckey, S.~Wang and H.~A. Carlson, \emph{Journal of Chemical
  Information and Modeling}, 2011, \textbf{51}, 2036--2046\relax
\mciteBstWouldAddEndPuncttrue
\mciteSetBstMidEndSepPunct{\mcitedefaultmidpunct}
{\mcitedefaultendpunct}{\mcitedefaultseppunct}\relax
\EndOfBibitem
\bibitem[Maia \emph{et~al.}(2020)Maia, Assis, de~Oliveira, da~Silva, and
  Taranto]{10.3389/fchem.2020.00343}
E.~H.~B. Maia, L.~C. Assis, T.~A. de~Oliveira, A.~M. da~Silva and A.~G.
  Taranto, \emph{Frontiers in Chemistry}, 2020, \textbf{8}, 343\relax
\mciteBstWouldAddEndPuncttrue
\mciteSetBstMidEndSepPunct{\mcitedefaultmidpunct}
{\mcitedefaultendpunct}{\mcitedefaultseppunct}\relax
\EndOfBibitem
\bibitem[Sink \emph{et~al.}(2010)Sink, Gobec, Pecar, and Zega]{Gobec2010}
R.~Sink, S.~Gobec, S.~Pecar and A.~Zega, \emph{Current Medicinal Chemistry},
  2010, \textbf{17}, 4231--4255\relax
\mciteBstWouldAddEndPuncttrue
\mciteSetBstMidEndSepPunct{\mcitedefaultmidpunct}
{\mcitedefaultendpunct}{\mcitedefaultseppunct}\relax
\EndOfBibitem
\bibitem[Gal and Ghahramani(2016)]{pmlr-v48-gal16}
Y.~Gal and Z.~Ghahramani, Proceedings of The 33rd International Conference on
  Machine Learning, New York, New York, USA, 2016\relax
\mciteBstWouldAddEndPuncttrue
\mciteSetBstMidEndSepPunct{\mcitedefaultmidpunct}
{\mcitedefaultendpunct}{\mcitedefaultseppunct}\relax
\EndOfBibitem
\bibitem[Scalia \emph{et~al.}(2020)Scalia, Grambow, Pernici, Li, and
  Green]{Scalia2020}
G.~Scalia, C.~A. Grambow, B.~Pernici, Y.-P. Li and W.~H. Green, \emph{Journal
  of Chemical Information and Modeling}, 2020,  2697--2717\relax
\mciteBstWouldAddEndPuncttrue
\mciteSetBstMidEndSepPunct{\mcitedefaultmidpunct}
{\mcitedefaultendpunct}{\mcitedefaultseppunct}\relax
\EndOfBibitem
\bibitem[Imrie \emph{et~al.}(2020)Imrie, Bradley, van~der Schaar, and
  Deane]{imrie2020deep}
F.~Imrie, A.~R. Bradley, M.~van~der Schaar and C.~M. Deane, \emph{Journal of
  chemical information and modeling}, 2020, \textbf{60}, 1983--1995\relax
\mciteBstWouldAddEndPuncttrue
\mciteSetBstMidEndSepPunct{\mcitedefaultmidpunct}
{\mcitedefaultendpunct}{\mcitedefaultseppunct}\relax
\EndOfBibitem
\bibitem[Geschwindner and Ulander(2019)]{Ulander2019}
S.~Geschwindner and J.~Ulander, \emph{Expert Opinion on Drug Discovery}, 2019,
  \textbf{14}, 1221--1225\relax
\mciteBstWouldAddEndPuncttrue
\mciteSetBstMidEndSepPunct{\mcitedefaultmidpunct}
{\mcitedefaultendpunct}{\mcitedefaultseppunct}\relax
\EndOfBibitem
\end{mcitethebibliography}
\bibliographystyle{rsc} 

\end{document}